\documentclass[twocolumn]{aastex63}
\usepackage{graphicx}
\usepackage{amsmath}
\usepackage{color}

\DeclareMathAlphabet{\mathsc}{OT1}{cmr}{m}{sc}
\def\testbx{bx}%
\DeclareRobustCommand{\ion}[2]{%
\relax\ifmmode
\ifx\testbx\f@series
{\mathbf{#1\,\mathsc{#2}}}\else
{\mathrm{#1\,\mathsc{#2}}}\fi
\else\textup{#1\,{\mdseries\textsc{#2}}}%
\fi}

\begin{document}

\title{Observational Constraints on the Physical Properties of Interstellar Dust in the Post-{\it Planck} Era}

\author[0000-0001-7449-4638]{Brandon S. Hensley}
\email{bhensley@astro.princeton.edu}
\affiliation{Department of Astrophysical Sciences,  Princeton
  University, Princeton, NJ 08544, USA}
\affiliation{Spitzer Fellow}

\author[0000-0002-0846-936X]{B. T. Draine}
\affiliation{Department of Astrophysical Sciences,  Princeton
  University, Princeton, NJ 08544, USA}

\date{\today}

\begin{abstract}
We present a synthesis of the astronomical observations constraining the wavelength-dependent extinction, emission, and polarization from interstellar dust from UV to microwave wavelengths on diffuse Galactic sightlines. Representative solid phase abundances for those sightlines are also derived. Given the sensitive new observations of polarized dust emission provided by the {\it Planck} satellite, we place particular emphasis on dust polarimetry, including continuum polarized extinction, polarization in the carbonaceous and silicate spectroscopic features, the wavelength-dependent polarization fraction of the dust emission, and the connection between optical polarized extinction and far-infrared polarized emission. Together, these constitute a set of constraints that should be reproduced by models of dust in the diffuse interstellar medium. 
\end{abstract}


\section{Introduction}
Interstellar dust is manifest at nearly all wavelengths of astronomical interest, scattering, absorbing, and emitting radiation from X-ray to radio wavelengths. Embedded in this diversity of phenomena are clues to the nature of interstellar grains---their size, shape, composition, and optical properties.

A combination of astronomical observations, laboratory studies, and theoretical calculations has informed a picture of interstellar dust that consists of, at minimum, amorphous silicate and carbonaceous materials \citep[see][for a review]{Draine_2003}. However, many questions remain as to the details of these components, e.g., their optical properties, porosity, purity, size distributions, shapes, and alignment, including whether the silicate and carbonaceous materials exist as distinct components or whether they are typically found in the same interstellar grains.

The astronomical data which constrain models of interstellar dust are extensive and ever increasing in detail. Determinations of solid phase abundances define the elemental makeup and mass of interstellar dust grains per H atom. Interstellar extinction has been measured from the far-UV (FUV) through the mid-infrared (MIR), including a number of spectral features suggesting specific materials. Emission from dust grains heated by the ambient interstellar radiation field has been observed from the near-infrared (NIR) through the microwave. Additionally, anomalous microwave emission (AME), thought to arise from rapidly rotating ultrasmall grains, is seen at radio frequencies while extended red emission (ERE), attributed to fluorescence, is observed in the optical. Polarization has been detected in both extinction and emission, including in some spectral features, placing additional constraints on the shapes, compositions, and alignment properties of interstellar grains.

With high-sensitivity far-infrared (FIR) imaging and polarimetry, the {\it Planck} satellite measured the properties of submillimeter polarized dust emission in unprecedented detail \citep{Planck_Int_XIX}. The very high submillimeter polarization fractions and the observed characteristic ratios between polarized FIR emission and polarized extinction at optical wavelengths have posed serious challenges to pre-{\it Planck} dust models \citep{Planck_Int_XIX, Planck_2018_XII}. It is imperative that these new findings guide the development of the next generation of models.

When presenting a new dust model, it has become customary to detail the set of observations that constrain it \citep[e.g.,][]{Mathis+Rumpl+Nordsieck_1977,Draine+Lee_1984,Zubko+Dwek+Arendt_2004,Draine+Fraisse_2009,Compiegne+etal_2011,Siebenmorgen+Voshchinnikov+Bagnulo_2014,Guillet+etal_2018}. Given the now vast array of observations that can be employed in calibrating and testing models, and given also the heterogeneity of the observations in terms of wavelengths covered and region observed, synthesizing a coherent set of model constraints can be as challenging as construction of the model itself. Sight line to sight line variations exist, and we can aim only to identify observational constraints that appear to be representative of characteristic environments. It is therefore the goal of this work to summarize the current state of observations constraining the properties of dust in the diffuse interstellar medium (ISM) and to establish a set of benchmark constraints against which models of interstellar dust can be tested. \added{We focus here on diffuse atomic sightlines where the observational constraints are most complete. As such, the properties of dust in dense clouds, such as ice features, will not be discussed.}

This paper is organized as follows: we first derive the solid phase abundances of the primary elemental constituents of dust in Section~\ref{sec:abundances}; then, we combine various observational data to derive the wavelength dependence of dust extinction (Section~\ref{sec:extinction}), polarized extinction (Section~\ref{sec:extpol}), emission (Section~\ref{sec:irem}), and polarized emission (Section~\ref{sec:ir_pol}) for a typical diffuse, high-latitude sightline. Finally, we present a summary of these constraints in Section~\ref{sec:summary}.

\section{Abundances}
\label{sec:abundances}

\begin{deluxetable}{ccll}
  \tablewidth{0pc}
      \tablecaption{Interstellar Abundances of Selected Elements\label{table:abundances}}
    \tablehead{\colhead{Element} & \colhead{$A$ [ppm]}
      & \colhead{Method} & \colhead{Reference(s)}}
    \startdata
    C & $324\pm38$ & Solar + Solar Twins &  2, 6, 11\\
    & $331\pm38$ & Solar + GCE Model & 2, 3, 6  \\
    & $358\pm82$ & Young F \& G Stars & 1 \\
    & $245\pm62$ & Young F \& G Stars & 5, 7\\
    & $190\pm77$ & B Stars & 1 \\
    & $214\pm20$ & B Stars & 8 \\
    O & $682\pm79$ & Solar + Solar Twins & 2, 6, 11 \\
    & $575\pm66$ & Solar + GCE Model & 2, 3, 6 \\
    & $445\pm156$ & Young F \& G Stars & 1 \\
    & $589\pm176$ & Young F \& G Stars & 4, 7 \\
    & $350\pm133$ & B Stars & 1 \\
    & $575\pm66$ & B Stars & 8 \\
    Mg & $52.9\pm4.9$ & Solar + Solar Twins & 2, 9, 11 \\
    & $45.7\pm4.2$ & Solar + GCE Model & 2, 3, 9 \\
    & $42.7\pm17.2$ & Young F \& G Stars & 1 \\
    & $44\pm21$ & Young F \& G Stars & 4, 7 \\
    & $23\pm7.0$ & B Stars & 1 \\
    & $36.3\pm4.2$ & B Stars & 8 \\
    Al & $3.5\pm0.3$ & Solar + Solar Twins & 2, 9, 11 \\
    & $3.5\pm1.8$ & Young F \& G Stars & 4, 7 \\
    Si & $44.6\pm3.1$ & Solar + Solar Twins & 5, 8 \\
    & $41.7\pm2.9$ & Solar + GCE Model & 2, 3, 9 \\
    & $39.9\pm13.1$ & Young F \& G Stars & 1 \\
    & $41\pm22$ & Young F \& G Stars & 4, 7 \\
    & $18.8\pm8.9$ & B Stars & 1 \\
    & $31.6\pm3.6$ & B Stars & 8 \\
    S & $17.2\pm1.2$ & Solar + Solar Twins & 2, 9, 11 \\ 
      & $17.4\pm1.2$ & Solar + GCE Model & 2, 3, 10 \\
    & $19.5\pm4.5$ & Young F \& G Stars & 4, 7 \\
    Ca & $3.2\pm0.2$ & Solar + Solar Twins & 2, 9, 11 \\
    & $3.0\pm2.7$ & Young F \& G Stars & 4, 7 \\
    Fe & $43.7\pm4.0$ & Solar + GCE Model & 2, 3, 10 \\
    & $27.9\pm7.7$ & Young F \& G Stars & 1 \\
    & $40.7\pm23.5$ & Young F \& G Stars & 4, 7 \\
    & $28.5\pm18.0$ & B Stars & 1 \\
    & $33.1\pm2.3$ & B Stars & 8 \\
    Ni & $2.1\pm0.2$ & Solar + Solar Twins & 2, 10, 11 \\
    & $1.8\pm0.3$ & Young F \& G Stars & 4, 7
    \enddata
    \tablenotetext{}{$^1$\citet{Sofia+Meyer_2001},
      $^2$\citet{Turcotte+Wimmer-Schweingruber_2002},
      $^3$\citet{Chiappini+Romano+Matteucci_2003},
      $^4$\citet{Bensby+etal_2005}, $^5$\citet{Bensby+Feltzing+2006},
      $^6$\citet{Asplund+etal_2009}, $^7$\citet{Lodders+etal_2009},
      $^8$\citet{Nieva+Przybilla_2012}, $^9$\citet{Scott+etal_2015a},
      $^{10}$\citet{Scott+etal_2015b}, $^{11}$\citet{Bedell+etal_2018}}
    \tablecomments{Abundances of selected elements derived from solar
      abundances (Refs. 6, 9, 10) corrected for diffusion (Ref. 2) and
      chemical enrichment (Refs. 3 and 11), from young F and
    G stars (Refs. 4, 5, and 7), and from B stars (Refs. 1 and 8).}
\end{deluxetable}

\begin{deluxetable}{cccc}
  \tablewidth{0pc}
      \tablecaption{Adopted Gas and Solid Phase Abundances of Selected
        Elements\label{table:solid_abundance}}
    \tablehead{\colhead{X} & \colhead{(X/H)$_{\rm ISM}$}
      & \colhead{(X/H)$_{\rm gas}$} & \colhead{(X/H)$_{\rm dust}$} \\
      & [ppm] & [ppm] & [ppm] }
    \startdata
    C & $324$ & $198$ & $126\pm56$ \\
    O & $682$ & $434$ & $249\pm94$ \\
    Mg & $52.9$ & $7.1$ & $45.8\pm4.9$ \\
    Al & $3.5$ & $0.1$ & $3.4\pm0.3$ \\
    Si & $44.6$ & $6.6$ & $38.0\pm3.1$ \\
    S  & $17.2$ & $9.6$ & $7.6\pm2.0$ \\
    Ca & $3.2$ & $0.1$ & $3.2\pm0.2$ \\
    Fe & $43.7$ & $0.88$ & $42.8\pm4.0$ \\
    Ni & $2.1$ & $0.04$ & $2.0\pm0.2$
    \enddata
    \tablecomments{ISM abundances based on Solar abundances
      \citep{Asplund+etal_2009, Scott+etal_2015a, Scott+etal_2015b} corrected for diffusion \citep{Turcotte+Wimmer-Schweingruber_2002} and with GCE \added{as determined from observations of young F and G stars} \citep{Bedell+etal_2018}. Gas phase abundances taken from \citet{Jenkins_2009} assuming moderate depletion (i.e., $F_* = 0.5$).}
\end{deluxetable}

The heavy elements that make up the bulk of the mass of grains are produced in stars which return material to the ISM via winds or ejecta. Some of the atoms remain in the gas while a fraction get locked in grains. Comparison of stellar and gas phase abundances of metals is thus an important observational constraint on grain models.

The elements C, O, Mg, Si, and Fe are depleted in the gas phase and compose most of the interstellar dust mass. In addition, Al, S, Ca, and Ni are also depleted and constitute a minor but non-negligible fraction of the dust mass. A dust model should account for the observed depletions of each of these elements. Other elements (e.g., Ti) are also present in the grains, but collectively account for $<1\%$ of the grain mass, and will not be discussed here.

While gas phase abundances are determined directly from absorption line spectroscopy, inferring the solid phase abundances from these measurements requires determination of the {\it total} abundance of
each element in the ISM. This is often done starting from the well-constrained Solar abundances and applying a correction for Galactic chemical enrichment (GCE) during the $\sim$4.6\,Gyr since the formation of the Sun.

Detailed 3D hydrodynamical modeling of the Solar atmosphere has yielded photospheric abundances of $\log\epsilon_{\rm C} = 8.43\pm0.05$,  $\log\epsilon_{\rm O} = 8.69\pm0.05$, $\log\epsilon_{\rm Mg} = 7.59\pm0.04$, $\log\epsilon_{\rm Al} = 6.43\pm0.04$, $\log\epsilon_{\rm Si} = 7.51\pm0.03$, $\log\epsilon_{\rm S} = 7.12\pm0.03$, $\log\epsilon_{\rm Ca} = 6.32\pm0.03$, $\log\epsilon_{\rm Fe} = 7.47\pm0.04$, and $\log\epsilon_{\rm Ni} = 6.20\pm0.04$ \citep{Asplund+etal_2009, Scott+etal_2015a, Scott+etal_2015b}, where

\begin{equation}
\log \epsilon_{\rm X} \equiv 12 + \log_{10}\left({\rm X}/{\rm H}\right) 
\end{equation}
and $\left({\rm X}/{\rm H}\right)$ is the number of atoms of element X per H atom. To convert these present-day photospheric abundances to protosolar abundances, we apply a diffusion correction of +0.03\,dex \citep{Turcotte+Wimmer-Schweingruber_2002}. We adopt these values as our reference protosolar abundances. 

The protosolar values are presumed to reflect the abundances in the ISM at the time of the Sun's formation 4.6\,Gyr ago. Present-day interstellar metal abundances are likely enhanced relative to these protosolar values. The chemical evolution model of \citet[][Model 7]{Chiappini+Romano+Matteucci_2003} predicts the C, O, Mg, Si, S, and Fe abundances to be enriched by 0.06, 0.04, 0.04, 0.08, 0.09, and 0.14 dex, respectively, relative to the protosolar values. \citet{Bedell+etal_2018} estimated the chemical enrichment as a function of time by determining the elemental abundances in Solar twins of various ages. If we assume $\Delta$[Fe/H] = 0.14 \citep{Chiappini+Romano+Matteucci_2003}, their results imply present-day enrichments of 0.05, 0.11, 0.10, 0.08, 0.11, 0.09, 0.16, and 0.09 dex for C, O, Mg, Al, Si, S, Ca, Ni respectively, where in the case of C, we have taken the weighted mean of the determinations based on \ion{C}{i} and CH. These results are summarized in  Table~\ref{table:abundances}. We apply the latter values to our reference protostellar abundances to define our reference ISM abundances, listed in the \replaced{first}{second} column of Table~\ref{table:solid_abundance}.

Interstellar abundances can also be inferred from observations of young stars. Studies of young ($<$ 1 Gyr) F and G stars \citep{Bensby+etal_2005, Bensby+Feltzing+2006} have yielded fairly concordant numbers for O, Mg, Al, Si, S, Ca, Fe, and Ni \citep[see][for review]{Lodders+etal_2009}. The C abundance, however, appears somewhat lower than would be predicted from the solar abundances. On the other hand, \citet{Sofia+Meyer_2001} report C, O, Mg, Si, and Fe abundances obtained from young ($\leq 2$\,Gyr) F and G stars that are in good agreement with the protosolar abundances plus enrichment, including the C abundance.

Photospheric abundances have also been determined for B stars with mostly consistent results, as summarized in Table~\ref{table:abundances}. However, the Si abundances determined from B stars are somewhat lower, with reported values of $18.8\pm8.9$\,ppm \citep{Sofia+Meyer_2001} and $31.6\pm2.6$\,ppm \citep{Nieva+Przybilla_2012}. Likewise, the Fe abundances are lower than those based on solar abundances by $\sim$10\,ppm.

Different determinations of the interstellar metal abundances are not yet fully concordant, and  the uncertainties quoted by any study using a specific class of objects may under-represent the underlying systematic uncertainties particular to that method. For the purposes of this work, we adopt abundances based on solar abundances plus enrichment as representative.

Once the baseline interstellar abundances have been determined, absorption line spectroscopy can be employed to determine the quantity of each element missing from the gas phase due to incorporation into grains. Compiling data over a large number of sightlines and gas species, \citet{Jenkins_2009} defined a parameter $F_*$ that quantifies the level of depletion of all metals along that sightline. $F_* = 0.5$, roughly the median depletion in the \citet{Jenkins_2009} sample, corresponded to sightlines with mean $n_{\rm H} \simeq 0.3$\,cm$^{-3}$, appropriate for diffuse \ion{H}{i}. Therefore, we adopt the gas phase abundances for $F_* = 0.5$ as representative for the diffuse sightlines of interest in this work.

In Table~\ref{table:solid_abundance}, we list the gas phase abundances of C, O, Mg, Si, S, Fe, and Ni corresponding to $F_* = 0.5$ and the relations for each element derived by \citet{Jenkins_2009}. For Al and Ca, we assume the level of depletion is the same as for Fe.

With the ISM and gas phase abundances constrained, we take the difference to determine the solid phase abundances, which we list in Table~\ref{table:solid_abundance}. We estimate the error bars by adding in quadrature those from Table~\ref{table:abundances} and the errors on the gas phase abundances inferred from \citet{Jenkins_2009}. Models of interstellar dust should account for the solid phase abundances presented here to within the observational and modeling uncertainties.

\added{The relative solid phase abundances of the elements in Table~\ref{table:solid_abundance} typically have a moderate dependence on the level of depletion (i.e., $F_*$). However, the number of depleted O atoms relative to depletion of atoms like Si, Mg, and Fe is a strong function of the parameter $F_*$, which seems to require major changes in the composition of the grain material formed as depletion proceeds. We therefore caution that the solid phase O abundance reported in Table~\ref{table:solid_abundance} for our idealized average diffuse sightline is especially uncertain. The O abundance poses a significant challenge for dust models. Models that assume solid-phase O to reside primarily in silicate material and metal oxides have difficulty accounting for the quantity of oxygen that is missing from the gas phase \citep{Jenkins_2009,Whittet_2010,Poteet+Whittet+Draine_2015}.}

\section{Extinction}
\label{sec:extinction}

\subsection{Introduction}
Interstellar dust attenuates light through both scattering and absorption. ``Extinction'' refers to the sum of these processes, and the wavelength dependence of interstellar extinction forms a key constraint on the properties of interstellar grains. Because interstellar dust preferentially extinguishes shorter wavelengths in the optical, the effects of extinction are often referred to as ``reddening.''

Extinction is typically measured in one of two ways. In the ``pair method,'' the spectrum of a reddened star is compared to an intrinsic spectrum derived from a set of standard unreddened stars \citep[e.g.,][]{Trumpler_1930, Bless+Savage_1972}. Alternatively, the stellar spectrum and the interstellar extinction can be modeled simultaneously with the aid of theoretical stellar spectra \citep[e.g.,][]{Fitzpatrick+Massa_2005,Schultheis+etal_2014,Fitzpatrick+etal_2019}. 

However, neither method readily yields the total extinction $A_\lambda$, where

\begin{equation}
\label{eq:a_lambda}
    A_\lambda \equiv 2.5 \log_{10} \frac{F^{\rm int}_\lambda}{F^{\rm obs}_\lambda}
    ~~~,
\end{equation}
$F_\lambda^{\rm int}=L_\lambda/4\pi D^2$ is the intrinsic (i.e., unreddened) flux, and $F^{\rm obs}_\lambda$ is the observed flux. However, if the wavelength dependence of the luminosity $L_\lambda$ is presumed to be known, the differential extinction for $\lambda^{-1}$ between two wavelengths is independent of distance $D$. Most empirical extinction curves are thus expressed as the ``selective'' extinction relative to a reference bandpass or wavelength $\lambda_0$ and written as

\begin{equation}
    E(\lambda-\lambda_0) \equiv A_\lambda - A_{\lambda_0}
    ~~~.
\end{equation}
To remove the dependence on the dust column, this is often then normalized to selective extinction between two reference bandpasses or wavelengths, classically the Johnson $B$ and $V$ bands, e.g., $E(\lambda-V)/E(B-V)$. The quantity 

\begin{equation}
R_V \equiv \frac{A_V}{E(B-V)}    
\end{equation}
is commonly used to parameterize the shape of the extinction curve.

As noted by many authors \citep[e.g.,][]{Blanco_1957,MaizApellaniz+etal_2014,Fitzpatrick+etal_2019}, the use of bandpasses rather than monochromatic wavelengths to normalize extinction curves becomes problematic at high precision because the measured extinction in a finite bandpass depends not just on the interstellar extinction law but also on the intrinsic spectrum of the object. We therefore focus where possible in this work on spectroscopic or spectrophotometric determinations of the interstellar extinction law.

Because we are principally interested in connecting observations to the properties of interstellar grains, we express our synthesized representative extinction law in terms of optical depth $\tau_\lambda$ rather than $A_\lambda$, which are related by

\begin{equation}
    \tau_\lambda \equiv \ln\left(\frac{F_\lambda^{\rm int}}{F_\lambda^{\rm obs}}\right) = \frac{A_\lambda}{2.5\log_{10} e} = \frac{A_\lambda}{1.0857}
    ~~~.
\end{equation}

\subsection{X-Ray Extinction}
\label{subsec:ext_xray}

Although measurement of absolute extinction is usually not possible at X-ray energies, the differential extinction associated with X-ray absorption features can be determined spectroscopically. Such spectroscopic measurements have been made across the O K edge at 530\,eV \citep[e.g.,][]{Takei+etal_2002}, Fe L edge at $\sim700$--$750$\,eV \citep[e.g.,][]{Paerels+etal_2001,Lee+etal_2009}, Mg K edge at 1.3\,keV \citep[e.g.,][]{Rogantini+etal_2020}, and Si K edge at 1.84\,keV \citep{Schulz+etal_2016, Zeegers+etal_2017, Rogantini+etal_2020}. {\it Chandra} and {\it XMM-Newton} both have sufficient spectral resolution to distinguish gas-phase absorption from extinction contributed by dust.

X-ray spectra have been interpreted as showing that interstellar silicates are Mg-rich \citep{Costantini+etal_2012, Rogantini+etal_2019}, and \citet{Westphal+etal_2019} conclude that most of the Fe is in metallic form. While the absorption profile of the $10\,\mu$m silicate feature has been interpreted as giving a $2\%$ upper limit on the crystalline fraction (see Section~\ref{subsec:ext_sil_features}), X-ray observations of the Mg and Si K edges have been interpreted as showing that $11-15\%$ of the silicate material is crystalline, \citep{Rogantini+etal_2019,Rogantini+etal_2020}.

Efforts to identify the specific minerals hosting the solid-phase C, O, Mg, Si, and Fe remain inconclusive because of not-quite-sufficient spectral resolution, limited signal-to-noise, and limited laboratory data. Scattering contributes significantly to the extinction \citep{Draine_2003b}, and therefore model comparisons depend not only on the composition of the dust, but also on the size and shape of the grains \citep{Hoffman+Draine_2016}. Future measurements of X-ray extinction and X-ray scattering (see Section~\ref{subsubsec:xray_sca}) offer the prospect of mineralogical identification. The key will be to interpret the observations using dust models together with all available observational constraints.

\subsection{UV Extinction}
\label{subsec:ext_uv}

Spectroscopy from the {\it International Ultraviolet Explorer} (IUE) has been one of the primary datasets for characterizing the interstellar extinction law in the UV since the 1980s \citep[e.g.,][]{Witt+etal_1984,Fitzpatrick+Massa_1986,Fitzpatrick+Massa_1988,Cardelli+Clayton+Mathis_1989,Valencic+Clayton+Gordon_2004}. Other notable measurements of UV extinction have been made by the {\it Copernicus} satellite \citep[e.g.][]{Cardelli+Clayton+Mathis_1989}, the {\it Orbiting and Retrievable Far and Extreme Ultraviolet Spectrometer} \citep[ORFEUS,][]{Sasseen+etal_2002}, the {\it Hubble Space Telescope} \citep[{\it HST}; e.g.,][]{Clayton+etal_2003}, and the {\it Far Ultraviolet Spectroscopic Explorer} \citep[{\it FUSE}; e.g.,][]{Gordon+etal_2009}. Extinction in the UV is characterized by a steep rise to short wavelengths, a prominent broad spectral feature at 2175\,\AA\ (see Section~\ref{subsubsec:2175}), and a notable lack of other substructure \citep{Clayton+etal_2003,Gordon+etal_2009}. 

\begin{figure*}
    \centering
        \includegraphics[width=\textwidth]{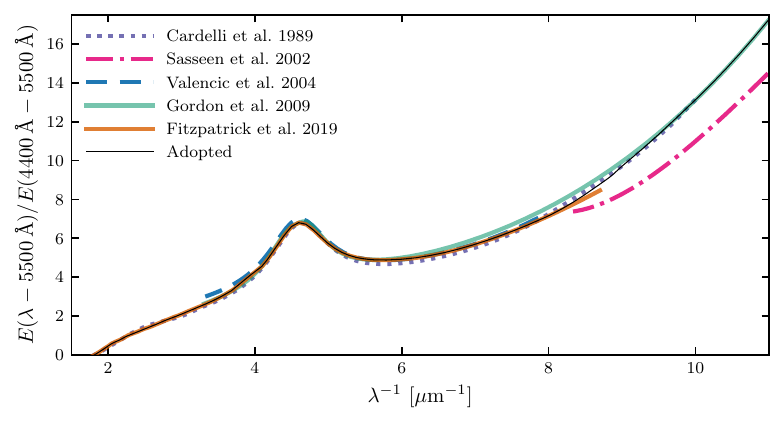}
    \caption{We present various determinations of the UV extinction curve of the diffuse Galactic ISM. The extinction law of \citet{Fitzpatrick+etal_2019} (orange solid) was derived in terms of monochromatic reddenings. However, those of \citet{Cardelli+Clayton+Mathis_1989} (purple dotted), \citet{Sasseen+etal_2002} (magenta dot-dashed), \citet{Valencic+Clayton+Gordon_2004} (blue dashed), and \citet{Gordon+etal_2009} (green solid) were derived with respect to finite bandpasses, i.e., $E(\lambda-V)/E(B-V)$. As discussed in the text, we present these curves by simply assuming perfect correspondence with $E\left(\lambda-5500\,\text{\AA}\right)/E\left(5500\,\text{\AA}-6410\,\text{\AA}\right)$.} \label{fig:ext_uv}
\end{figure*}

Spectroscopic characterization of interstellar extinction from UV to optical wavelengths was recently undertaken by \citet{Fitzpatrick+etal_2019}, who used {\it HST} Space Telescope Imaging Spectrograph (STIS) spectroscopy extending from 290-1027\,nm to complement {\it IUE} UV data. Additionally, JHK photometry from the Two-Micron All Sky Survey (2MASS) was used to extend the analysis into the near-infrared. On the basis of these data toward a curated sample of 72 O and B stars, they derived a mean extinction law having $A( 5500\,\text{\AA})/E(4400\,\text{\AA}-5500\,\text{\AA}) = 3.02$, corresponding approximately to $R_V = 3.1$. Because of the narrow-band observations, the resulting extinction curve is monochromatic and normalized using the extinction at 4400 and 5500\,\AA\ rather than the Johnson $B$ and $V$ bands, respectively. We illustrate this curve in Figures~\ref{fig:ext_uv} and \ref{fig:ext_op}.

On the basis of UV, optical, and NIR data toward a sample of 45 stars studied in the UV by \citet{Fitzpatrick+Massa_1988}, \citet{Cardelli+Clayton+Mathis_1988} presented an analytic parameterization for the extinction between 3.3 and 8\,$\mu$m$^{-1}$ as a function of $R_V$. This law was extended to the range 0.3 to 10\,$\mu$m$^{-1}$ by \citet{Cardelli+Clayton+Mathis_1989}. Combining {\it IUE} spectroscopy and 2MASS data along 417 lines of sight, \citet{Valencic+Clayton+Gordon_2004} further refined this parameterization in the 3.3 to 8.0 $\mu$m$^{-1}$ range\footnote{Note corrected numbers in \citet{Valencic_erratum}.}. We note, however, that the extinction law in this range was not formulated to join smoothly with the adjacent sections of the extinction law parameterized by \citet{Cardelli+Clayton+Mathis_1989}. Finally, \citet{Gordon+etal_2009} used the functional form\footnote{Note corrected numbers in \citet{Gordon2009_erratum}.} presented in \citet{Cardelli+Clayton+Mathis_1989} to fit 75 extinction curves measured with {\it FUSE} data from 3.3 to 11\,$\mu$m$^{-1}$.

We include the extinction laws of \citet{Cardelli+Clayton+Mathis_1989}, \citet{Valencic+Clayton+Gordon_2004}, and \citet{Gordon+etal_2009} in Figure~\ref{fig:ext_uv}. These extinction laws were derived in terms of $E(\lambda-V)/E(B-V)$ rather than monochromatic equivalents. Applying the correction factors to account for the finite bandpasses suggested by \citet{Fitzpatrick+etal_2019} (their Equation~4) results in curves that deviate more substantially from unity at 4400\,\AA\ and zero at 5500\,\AA\ than applying no correction. It is also the case that the $R_V = 3.1$ curve using the \citet{Cardelli+Clayton+Mathis_1989} parameterization does not precisely have $A_V/E(B-V) = 3.1$ \citep[see discussion in][]{MaizApellaniz_2013}. Given these issues, we simply assume that $E(B-V)$ corresponds exactly to $E(4400\,\text{\AA}\ -\ 5500\,\text{\AA})$ to convert the $R_V = 3.1$ curves to monochromatic reddenings.

\citet{Sasseen+etal_2002} made a determination of the mean FUV (910--1200\,\AA) extinction law using observations of eleven pairs of B stars with the {\it ORFEUS} spectrometer. This curve is also plotted in Figure~\ref{fig:ext_uv} where, as with several of the other curves presented, we do not apply any corrections to translate from the reported $E(\lambda-V)/E(B-V)$ to monochromatic reddenings. While the shape of this curve is in general agreement with that of \citet{Cardelli+Clayton+Mathis_1989} and \citet{Gordon+etal_2009}, there is significantly less FUV extinction per $E(B-V)$.

As Figure~\ref{fig:ext_uv} illustrates, there is general agreement among extinction curves in the UV. The \citet{Fitzpatrick+etal_2019} and \citet{Gordon+etal_2009} curves correspond closely between 3.3 and 6\,$\mu$m$^{-1}$, while that of \citet{Valencic+Clayton+Gordon_2004} agrees better with \citet{Fitzpatrick+etal_2019} between 5 and 8\,$\mu$m$^{-1}$. For our representative extinction curve, we therefore employ the \citet{Fitzpatrick+etal_2019} curve from 5500\,\AA\ to 8\,$\mu$m$^{-1}$, and then match onto the curve of \citet{Gordon+etal_2009} to extend to 11\,$\mu$m$^{-1}$. This is accomplished by using the \citet{Cardelli+Clayton+Mathis_1989} curve between 8 and 10\,$\mu$m$^{-1}$.

\subsection{Optical Extinction}
\label{subsec:ext_op}

\begin{figure}
    \centering
        \includegraphics[width=\columnwidth]{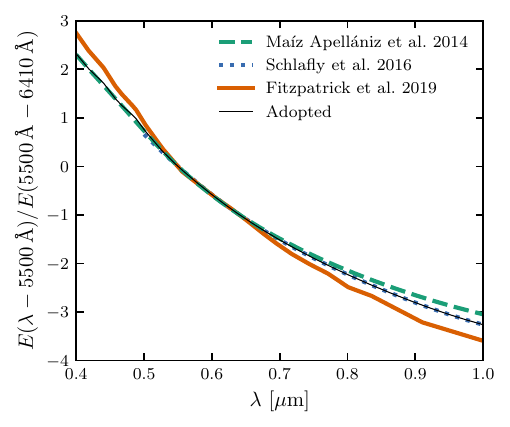}
    \caption{The extinction laws of \citet{MaizApellaniz+etal_2014} (green \deleted{dot-}dashed), \citet{Schlafly+etal_2016} (blue dotted), and \citet{Fitzpatrick+etal_2019} (orange solid) are compared at optical wavelengths. The normalization corresponds roughly to $E(V-R)$ which we adopt since the \citet{Schlafly+etal_2016} curve does not extend to wavelengths shorter than 500\,nm.} \label{fig:ext_op}
\end{figure}

While the extinction curve of the diffuse ISM has been well determined from UV to optical wavelengths over decades of observations, it is only recently that spectrophotometric observations have enabled detailed characterization at optical wavelengths. In this section, we compare determinations of the mean Galactic extinction curve from 500\,nm to 1\,$\mu$m.

In addition to \citet{Fitzpatrick+etal_2019}, a recent determination of the optical extinction law using spectroscopy is that of \citet{MaizApellaniz+etal_2014}, who used the Fibre Large Array Multi-Element Spectrograph (FLAMES) on the Very Large Telescope to determine the extinction toward 83 O and B stars in 30~Doradus. These spectroscopic data extended from 3960–-5071\,\AA\ and were supplemented with both 2MASS JHK and {\it HST} Wide Field Camera 3 photometry (UBVI and H$\alpha$) to test and revise the \citet{Cardelli+Clayton+Mathis_1989} extinction law from the optical to the near-infrared. Outside this wavelength range, \citet{MaizApellaniz+etal_2014} tied the extinction law to that of \citet{Cardelli+Clayton+Mathis_1989}. The resulting curve is presented in Figure~\ref{fig:ext_op} alongside that of \citet{Fitzpatrick+etal_2019}. While there is general consistency between the two extinction laws, there are also significant departures. As 30~Doradus is located in the Large Magellanic Cloud (LMC), the extinction has a contribution from the LMC dust which may differ systematically from that of the Galaxy.  We therefore seek comparisons with other observations.

\citet{Schlafly+etal_2016} determined the extinction toward 37,000 APOGEE stars in ten photometric bands from $g$ (503.2\,nm) to WISE 2 (4.48\,$\mu$m). This wavelength coverage does not extend far enough blueward to apply the normalization used in Figure~\ref{fig:ext_uv}, and indeed \citet{Schlafly+etal_2016} note that different methods of extrapolating their extinction law to the $B$ band yield $R_V$s that differ by a few tenths. Thus, Figure~\ref{fig:ext_op} presents a different comparison using $E(5500\,\text{\AA}-6410\,\text{\AA})$ as the normalization factor, roughly equivalent to $E(V-R)$. Because of the explicit treatment of the bandpasses, the \citet{Schlafly+etal_2016} extinction curve is defined with respect to monochromatic wavelengths.

From 500 to $\sim800$\,nm, the \citet{MaizApellaniz+etal_2014} and \citet{Schlafly+etal_2016} curves are in close agreement. We note that the \citet{MaizApellaniz+etal_2014} extinction law defaults to that of \citet{Cardelli+Clayton+Mathis_1989} at wavelengths longer than $\sim800$\,nm. Indeed, \citet{Schlafly+etal_2016} note that the \citet{Cardelli+Clayton+Mathis_1989} parameterization provides a poor fit to the infrared data for the full range of $R_V$ studied while the \citet{MaizApellaniz+etal_2014} law is an excellent fit in the optical.

\citet{Wang+Chen_2019} employed {\it Gaia} parallaxes for a sample of more than 61,000 red clump stars in APOGEE to overcome the distance/attenuation degeneracy and derive a mean interstellar extinction law in 21 photometric bands. When expressed as color excess ratios $E(\lambda-\lambda_1)/E(\lambda_2-\lambda_1)$, their mean curve agrees with that of \citet{Schlafly+etal_2016} to within a few percent over the full 0.5--4.5\,$\mu$m wavelength range.

Given these corroborating studies, we adopt the extinction law of \citet{Schlafly+etal_2016} from 550\,nm to the IR. However, converting from $E(\lambda-5500\,\text{\AA})/E(4400\,\text{\AA}-5500\,\text{\AA})$ to a quantity like $A_\lambda/A(5500\,\text{\AA})$ requires a measurement of the absolute extinction at some wavelength. This is because a single reddening law is consistent with a family of extinction laws that differ by an additive offset common to all wavelengths over which the reddening has been measured. The classic $R_V = 3.1$ is relatively well-determined from the fact that the infrared extinction is much smaller than the optical and UV extinction, and so measurement of reddening relative to a NIR band, e.g., $E(V-L)/E(B-V)$, constrains any component common to all bands sufficiently well, i.e., $E(V-L)/E(B-V) \approx R_V$. As determinations of the extinction curve are made at increasingly long wavelengths, the sensitivity to the size of this common component increases. We explore this issue in more detail in the following section.

\subsection{NIR Extinction}
\label{subsec:ext_nir}

\begin{figure}
    \centering
        \includegraphics[width=\columnwidth]{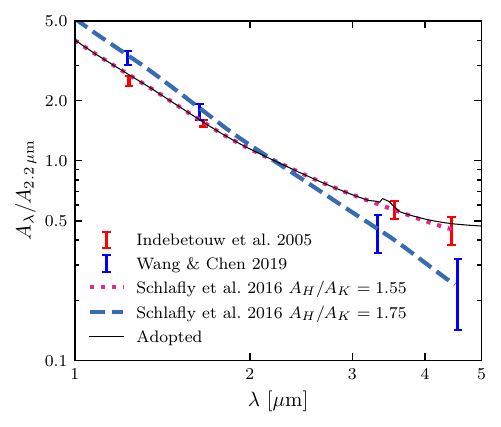}
    \caption{We compare the NIR/MIR broadband extinction curves of \citet{Indebetouw+etal_2005} along a diffuse sightline in the Galactic plane ($\ell = 42^\circ$) and \citet{Wang+Chen_2019} toward a sample of more than 61,000 red clump stars in the APOGEE survey. For both of these curves, we report extinction relative to the $K_s$ band. We compare these determinations to the \citet{Schlafly+etal_2016} reddening law assuming two values of $A_H/A_K$: 1.55 from \citet{Indebetouw+etal_2005} and 1.75, a representative average of recent measurements (see Section~\ref{subsec:ext_nir}).} \label{fig:ext_nir}
\end{figure}

The NIR extinction law from $\sim$1--5\,$\mu$m is often approximated as a power law $A_\lambda \propto \lambda^{-\alpha}$. A foundational analysis of NIR extinction was made by \citet{Rieke+Lebofsky_1985}, who measured extinction toward $o$\,Sco \citep[$A_V = 2.7$,][]{Whittet_1988b}, Cyg\,OB2-12 \citep[$A_V \simeq 10.2$,][]{Humphreys_1978,TorresDodgen+etal_1991}, and several heavily reddened sources toward the Galactic Center ($A_V$ between 23 and 35). The widely-used extinction law of \citet{Cardelli+Clayton+Mathis_1989} relies on the extinction curve determined by \citet{Rieke+Lebofsky_1985} at wavelengths longer than $J$ band ($\lambda_J \simeq 1.23\,\mu$m), employing $\alpha = 1.61$ for $0.91 < \lambda/\mu{\rm m} < 3.3$.

Many other early determinations of $\alpha$ likewise found values in the range $\sim$ 1.6--1.85 (see the reviews of \citet{Draine_1989} and \citet{Mathis_1990}). However, an analysis by \citet{Stead+Hoare_2009} demonstrated that the value of $\alpha$ derived from fits to extinction in the $JHK$ photometric bands depends sensitively on how the bandpasses are treated, particularly for highly reddened sources. Accounting explicitly for these bandpass effects in sources of different intrinsic spectra and levels of reddening, and using photometry from both the United Kingdom Infrared Deep Sky Survey (UKIDSS) and 2MASS, they recommend a mean value of $\alpha = 2.15\pm0.05$, significantly larger than most earlier determinations. Recently, a similar study using 2MASS photometry found $\alpha = 2.27$ with an uncertainty of $\sim1$\% \citep{MaizApellaniz2020}.

While the power law approximation is both simple and effective, \citet{Fitzpatrick+Massa_2009} demonstrated that extinction between the $I$ ($\lambda_I \simeq 0.798\,\mu$m)and $K_s$ ($\lambda_{K_s} \simeq 2.16\,\mu$m) bands is better represented by a modified power law in which $\alpha$ increases between 0.75 and 2.2\,$\mu$m. They proposed instead a function of the form

\begin{equation}
    \frac{A_\lambda}{E\left(B-V\right)} \propto \frac{1}{1 + \left(\lambda/\lambda_0\right)^\gamma}
    ~~~,
\end{equation}
with $\lambda_0 = 0.507$\,$\mu$m. The fit values of $\gamma$ varied considerably from sightline to sightline, ranging from $\sim1.8$--2.8, and the constant of proportionality was found to depend on $R_V$. \citet{Schlafly+etal_2016} found excellent agreement in the NIR between this parameterization with $\gamma \simeq 2.5$ and their mean extinction law. While this functional form captures flattening of the extinction law at the shortest wavelengths in this range, other studies have noted an apparent flattening of the NIR extinction law at the longest wavelengths as well, particularly in comparing the slope of the extinction curve between $J$ and $H$ ($\lambda_H \simeq 1.63\,\mu$m) to the slope between $H$ and $K_s$ \citep[e.g.,][]{Fritz+etal_2011,Hosek+etal_2018,NoguerasLara+etal_2020}. Such behavior is not unexpected given indications of a relatively flat MIR extinction curve (see Section~\ref{subsec:mir_ext}).

The assumption of a power law can have dramatic effect on the conversion from reddening to extinction. If $A_\lambda \propto \lambda^{-\alpha}$, then

\begin{equation}
\label{eq:ejhk_alpha}
    \frac{E\left(J-H\right)}{E\left(H-K\right)} = \frac{\left(\frac{\lambda_H}{\lambda_J}\right)^\alpha - 1}{1 - \left(\frac{\lambda_H}{\lambda_K}\right)^\alpha}
    ~~~.
\end{equation}
With a sample of 37,000 stars, \citet{Schlafly+etal_2016} made precise determinations of interstellar reddening in these bands. Inserting their measured reddenings into Equation~\ref{eq:ejhk_alpha} yields $\alpha = 2.30$. As discussed in Section~\replaced{\ref{subsec:ext_nir}}{\ref{subsec:ext_op}}, however, a single reddening law is consistent with a family of extinction laws that differ by an additive constant. One method of placing a limit on this constant is to require the extinction in the longest wavelength band to be positive. Another more constraining method is to find the additive constant such that the ratio of the extinction in two bands agrees with a measured value. We find that $\alpha \simeq 2.30$ between $J$ and $K$ can be achieved by employing the \citet{Schlafly+etal_2016} reddening law and imposing $A_H/A_K = 1.87$. However, this same reddening law is consistent with a (wavelength-dependent) logarithmic slope of $\sim1.7$ in the NIR when instead requiring $A_H/A_K = 1.55$ \citep[as determined by][]{Indebetouw+etal_2005}.

It is therefore unclear whether the large values of $\alpha$ found by \citet{Stead+Hoare_2009} and \citet{MaizApellaniz2020} are indeed more physical due to the more careful treatment of the bandpasses or whether they are biased toward higher values of $\alpha$ by forcing extinction in the $JHK$ bands to conform precisely to a power law. An independent constraint on the absolute extinction is needed to break this degeneracy.

The default curve put forward by \citet{Schlafly+etal_2016} employs $A_H/A_K = 1.55$ as determined by \citet{Indebetouw+etal_2005}. In that study, the absolute extinction was constrained along diffuse sight lines in the Galactic plane with $\ell = 42^\circ$ by measuring the extinction toward K giants, which are well-localized in color space, under the assumption that extinction per unit distance is constant in the Galactic plane. \citet{Wang+Chen_2019} used {\it Gaia} parallaxes to measure the reddening as a function of distance modulus toward a sample of more than 60,000 red clump stars. They found $A_H/A_{K_s} = 1.75$, noting agreement with \citet{Chen+etal_2018} who used 55 classical Cepheids to measure distance to the Galactic Center and derived $A_H/A_{K_s} = 1.717$. Photometry of red clump stars toward the Galactic Center has yielded relatively concordant values of $\sim1.69\pm0.03$ \citep{Nishiyama+etal_2006,Nagatomo+etal_2019}, $1.76\pm0.10$ \citep{Schodel+etal_2010}, and $1.84\pm0.03$ \citep{NoguerasLara+etal_2020}.

The steep NIR extinction laws implied by large values of $A_H/A_K$ are difficult to reconcile with relatively flat extinction between 4--8\,$\mu$m and comparisons between visual extinction and extinction in the 9.7\,$\mu$m feature, as we discuss in the next section. The NIR extinction is sensitive to relative abundance of the largest interstellar grains, and so sightlines passing through molecular gas, where grains grow to larger sizes through coagulation, may have systematically different properties. It is unclear if this effect is responsible for the discrepancy between the observations of \citet{Indebetouw+etal_2005} on a relatively diffuse sightline and those toward the Galactic Center.

Ultimately, on the basis of the observed properties of the MIR extinction, we adopt as our representative NIR extinction curve the reddening law of \citet{Schlafly+etal_2016} with $A_H/A_K = 1.55$ to convert to extinction. We present the resulting extinction law in Figure~\ref{fig:ext_nir}, where we compare it to the same reddening law derived assuming $A_H/A_K = 1.75$ instead. Further studies of NIR extinction along diffuse sightlines are needed to clarify the steepness of the interstellar extinction curve and its variations with the local environment.

\subsection{MIR Extinction}
\label{subsec:mir_ext}
\begin{figure*}
    \centering
        \includegraphics[width=\textwidth]{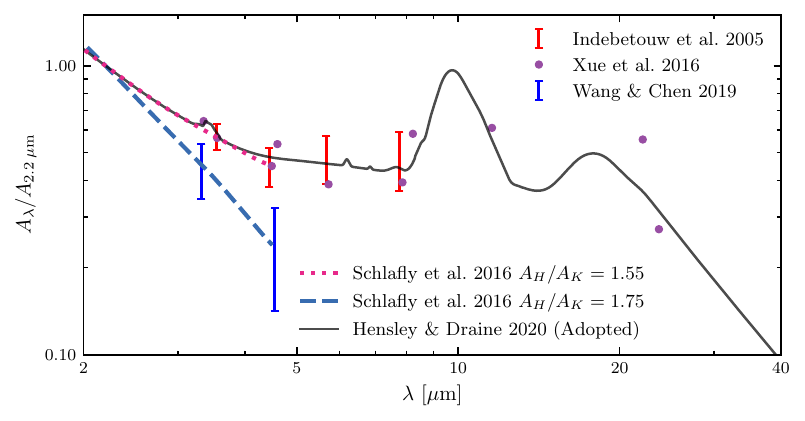}
    \caption{Various determinations of the MIR extinction law are presented, normalized to 2.2\,$\mu$m or the $K_s$ band as appropriate. Two versions of the \citet{Schlafly+etal_2016} curve are shown, one assuming $A_H/A_K = 1.55$ (dotted) and the other 1.75 (dashed). The extinction toward Cyg\,OB2-12 based on \replaced{ISO SWS}{{\it ISO}-SWS} and {\it Spitzer} IRS data is shown in black \citep{Hensley+Draine_2020} and is the basis of our synthesized extinction curve for $\lambda > 2.2\,\mu$m. Note that it matches onto the \citet{Schlafly+etal_2016} curve with $A_H/A_K = 1.55$ by design. We also present a few representative broadband determinations using combinations of {\it Spitzer} IRAC, {\it WISE}, and {\it AKARI} photometry, including \citet{Indebetouw+etal_2005} (red error bars), \citet{Xue+etal_2016} ($A_{K_s} < 0.5$ sample, purple circles), and \citet{Wang+Chen_2019} (blue error bars).} \label{fig:ext_mir}
\end{figure*}

The MIR extinction is dominated by continuum extinction between 3--8\,$\mu$m and by the 9.7 and 18\,$\mu$m silicate features longward of 8\,$\mu$m. We focus here on the former, deferring discussion of the silicate features to Section~\ref{subsec:ext_sil_features}. Carbonaceous MIR extinction features are discussed in Section~\ref{subsubsec:carbon_features_mir}. 

Some early determinations of the MIR extinction suggested a continuation of the NIR power law with a sharp minimum at 7\,$\mu$m \citep[e.g.,][]{Rieke+Lebofsky_1985,Bertoldi+etal_1999, Rosenthal+Bertoldi+Drapatz_2000, Hennebelle+etal_2001}. However, a growing body of work suggests that the MIR extinction is relatively flat between $\sim 4$ and 8\,$\mu$m across a diversity of sightlines and values of $R_V$.

Sightlines toward the Galactic Center have been well-measured in extinction and were the first to suggest, via observation of hydrogen recombination lines, a flattening of the extinction law in the MIR \citep{Lutz+etal_1996, Lutz_1999}. Subsequent broadband and spectroscopic observations toward the
Galactic center \citep{Nishiyama+etal_2006, Nishiyama+etal_2008, Nishiyama+etal_2009, Fritz+etal_2011} and the Galactic plane \citep{Jiang+etal_2003, Jiang+etal_2006, Gao+Jiang+Li_2009} have proven consistent with a relatively flat extinction law. Likewise, \citet{Flaherty+etal_2007} found good agreement with the \citet{Lutz+etal_1996} extinction curve when measuring the extinction toward nearby star-forming regions where the extinction was dominated by molecular gas. Observing in the dark cloud Barnard~59 ($A_K \sim 7$, $A_V \sim 59$), \citet{RomanZuniga+etal_2007} measured a 1.25--7.76\,$\mu$m extinction law consistent with that of \citet{Lutz+etal_1996}.

We seek the properties of dust in the diffuse ISM, which may be systematically different from these more heavily extinguished sightlines. However, the relatively flat extinction law between $\sim$3 and 8\,$\mu$m appears fairly universal. Combining {\it Spitzer} and 2MASS observations on an ``unremarkable'' region in the Galactic plane centered on $\ell = 42^\circ$, $b = 0.5^\circ$, \citet{Indebetouw+etal_2005} derived a extinction curve in agreement with \citet{Lutz+etal_1996}. \citet{Zasowski+etal_2009} derived an average extinction curve over 150$^\circ$ in the Galactic midplane also using {\it Spitzer} and 2MASS photometry, finding excellent agreement with \citet{Indebetouw+etal_2005}. Further, they note consistency between their result and extinction curves in low extinction regions in molecular clouds measured by \citet{Chapman+etal_2009}. \citet{Wang+etal_2013} measured the IR extinction law in regions of the Coalsack nebula that sampled a range of environments from diffuse to dark, finding a relatively universal shape of the MIR extinction across environments. \citet{Xue+etal_2016} derived a relatively flat MIR extinction curve toward a sample of G and K-giants in the {\it Spitzer} IRAC bands, in agreement with recent studies and sharply discrepant with a deep minimum in the extinction curve at $\sim7\,\mu$m.

The {\it Spitzer} Infrared Spectrograph (IRS) enables spectroscopic determination of the extinction law from $\sim5$--$37\,\mu$m. Employing IRS spectra toward a sample of five O and B stars, \citet{Shao+etal_2018} derived a relatively flat extinction curve between 5 and 7.5\,$\mu$m. Also using IRS data, \citet{Hensley+Draine_2020} determined a nearly identical extinction curve toward Cyg\,OB-12 in the 5--8\,$\mu$m range.

On the basis of these data, we conclude that a relatively flat extinction curve between $\sim4$--$8\,\mu$m is universal and typical of even of the diffuse ISM having $R_V \approx 3.1$, not just sightlines with large values of $R_V$. We summarize a selection of these observations in Figure~\ref{fig:ext_mir}. It must be cautioned, however, that the conversion from reddenings to extinction in many of these studies was accomplished by assuming a power law form in the NIR, and thus uncertainty still remains in both the precise shape and amount of 4--8\,$\mu$m extinction relative to the NIR.

To create a composite extinction law, we join the \citet{Schlafly+etal_2016} curve (with $A_H/A_K = 1.55)$ described in Section~\ref{subsec:ext_nir} to the extinction measured toward Cyg\,OB2-12 by \citet{Hensley+Draine_2020}. The latter study presented a synthesized extinction curve by joining the measured 6--37\,$\mu$m extinction inferred from {\it Spitzer} IRS measurements to the \citet{Schlafly+etal_2016} extinction law likewise assuming $A_H/A_K = 1.55$. As illustrated in Figure~\ref{fig:ext_mir}, this provides a good representation of other studies of extinction in the 4--8\,$\mu$m range. 

As discussed in Section~\ref{subsec:ext_nir}, $A_H/A_K = 1.55$ is low relative to several recent determinations, which favor a value of $\simeq1.75$. On the other hand, the \citet{Schlafly+etal_2016} extinction law having $A_H/A_K = 1.75$ shows no evidence for flattening even out to 4.5\,$\mu$m and implies lower 4--8\,$\mu$m extinction relative to $K$ band than inferred from a number of studies (see Figure~\ref{fig:ext_mir}). As we discuss in the following section, our adopted extinction curve has a value of $A_V/\Delta\tau_{9.7} = 20.0$, at the upper end of the observed range \citep[$\sim18.5\pm2$,][]{Draine_2003}. Joining a representative MIR extinction profile to an NIR extinction law with a higher value of $A_H/A_K$ would result in a larger $A_V/\Delta\tau_{9.7}$, exacerbating this tension. More work is needed to fully reconcile the existing observations of NIR and MIR extinction, and we thus present our synthesized curve as only our current best estimate of the true interstellar extinction.

Finally, we note that \citet{Schlafly+etal_2016} determined the interstellar extinction only in broad photometric bands and thus their resulting extinction curve does not contain spectral features. In contrast, \citet{Hensley+Draine_2020} used spectroscopic {\it ISO}-SWS data to determine the profile of the the 3.4\,$\mu$m spectroscopic feature toward Cyg\,OB2-12, which can be seen in Figure~\ref{fig:ext_mir}. We discuss this and other other spectroscopic features in greater detail in the following sections.

\subsection{Silicate Features}
\label{subsec:ext_sil_features}

\begin{figure}
    \centering
        \includegraphics[width=\columnwidth]{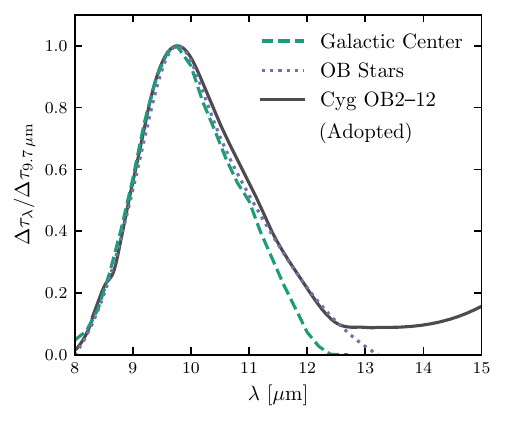}
    \caption{We compare three determinations of the profile of the 9.7\,$\mu$m silicate feature on different sightlines: toward the Galactic Center \citep{Kemper+Vriend+Tielens_2004}, a sample of five O and B stars \citep{Shao+etal_2018}, and Cyg\,OB2-12 \citep{Hensley+Draine_2020}. The agreement between these profiles suggests a universality of the silicate feature throughout the ISM. Residual differences between the profiles may be attributable to different treatments of the underlying continuum extinction.} \label{fig:ext_silprof} 
\end{figure}

In addition to smooth continuum extinction provided by the ensemble of interstellar dust grains, there are well-studied extinction features attributable to specific grain species. Prominent among these are features at 9.7 and 18\,$\mu$m that have been identified with silicate material, the former arising from the Si-O stretching mode and the latter from the O-Si-O bending mode.

The 9.7\,$\mu$m feature was discovered as a circumstellar emission feature \citep{Gillett+Low+Stein_1968, Woolf+Ney_1969}. \citet{Woolf+Ney_1969} demonstrated that the feature was consistent with the expected behavior of silicate material, a claim strengthened by the discovery of a second feature at 18\,$\mu$m \citep{Forrest+McCarthy+Houck_1979}. Subsequent observations have revealed that these features are not only found in circumstellar emission, but are also ubiquitous in absorption in the diffuse ISM \citep[see, e.g.][]{vanBreemen+etal_2011}.

The sightline to the Galactic Center has enabled detailed study of both the 9.7\,$\mu$m \citep{Roche+Aitken_1985, Smith+Aitken+Roche_1990, Kemper+Vriend+Tielens_2004} and 18\,$\mu$m features \citep{McCarthy+etal_1980} by virtue of its substantial dust column. \citet{Roche+Aitken_1985} found that the $V$ band extinction relative to the optical depth $\Delta\tau_{9.7}$ of the silicate feature at 9.7\,$\mu$m has a value of $A_V/\Delta\tau_{9.7} = 9\pm1$. \citet{Kemper+Vriend+Tielens_2004} employed {\it ISO} observations toward two carbon-rich Wolf-Rayet stars located toward the Galactic Center to derive the profile of the 9.7\,$\mu$m silicate feature $\Delta\tau_\lambda/\Delta\tau_{9.7\,\mu{\rm m}}$, which we plot in Figure~\ref{fig:ext_silprof}.

With heavy visual extinction \citep[$A_V \simeq 10.2$\,mag][]{Humphreys_1978,TorresDodgen+etal_1991} and yet a lack of ice features, the sightline toward the blue hypergiant Cyg\,OB2-12 is ideal for studying extinction arising from the diffuse atomic ISM \citep{Whittet_2015}. The 9.7\,$\mu$m silicate feature on this sightline was first observed by \citet{Rieke_1974}, and subsequent observations have produced detailed determinations of the both the 9.7 and 18\,$\mu$m silicate features \citep{Whittet+etal_1997,Fogerty+etal_2016,Hensley+Draine_2020}. In Figure~\ref{fig:ext_silprof}, we compare the Cyg\,OB2-12 feature profile determined by \citet{Hensley+Draine_2020} to that of the Galactic Center \citep{Kemper+Vriend+Tielens_2004} and a sample of O and B stars \citep{Shao+etal_2018}. 

The agreement between these profiles corroborates other studies noting a relatively universal silicate feature profile in the diffuse ISM \citep[e.g.,][]{Chiar+Tielens_2006,vanBreemen+etal_2011}. Interstellar dust models should therefore be compatible with this profile, which has FWHM $\simeq 2.2\,\mu$m. As noted by \citet{Chiar+Tielens_2006}, this average feature profile is narrower than the profile seen in emission toward the Trapezium region \citep[FWHM $\simeq 3.45\,\mu$m,][]{Forrest+Gillett+Stein_1975}, which was used to calibrate some models \citep[e.g.,][]{Draine+Lee_1984}.

Dust models should also be able to reproduce the observed strength of the feature. The extinction curve we synthesize in this work has $A_{5500\,\text{\AA}}/\Delta\tau_{9.7} = 20.0$. Comparing a variety of measurements toward Wolf-Rayet stars and toward Cyg\,OB2-12, \citet{Draine_2003} suggested a mean value $A_V/\Delta\tau_{9.7} = 18.5\pm2$, consistent with our composite curve.

Determination of the 18\,$\mu$m feature profile is made difficult by uncertainty in the underlying continuum extinction \citep[see discussion in][]{vanBreemen+etal_2011,Hensley+Draine_2020}. $\Delta\tau_{18}/\Delta\tau_{9.7}$ is typically found to be of order 0.5 \citep{Chiar+Tielens_2006,vanBreemen+etal_2011,Hensley+Draine_2020}. In performing model fits to the emission from \replaced{Cyg\,OB-2}{Cyg\,OB2-12} and its stellar wind, \citet{Hensley+Draine_2020} required that the extinction longward of 18\,$\mu$m extrapolate to values estimated from the FIR emission\added{. Further, this extrapolation was required to have} \deleted{with} a functional form \replaced{approximating the dust opacity law also inferred from FIR emission}{consistent with power law fits to the FIR emission}. Thus, while the 18\,$\mu$m feature itself is difficult to isolate from the total extinction, the long wavelength behavior of the extinction curve synthesized here is both physically and empirically motivated and serves as a reasonable best estimate.

Just as the presence of the 9.7 and 18\,$\mu$m silicate features constrains grain models, the {\it absence} of certain features likewise informs our understanding of the composition of interstellar dust. The 11.2\,$\mu$m feature arising from silicon carbide (SiC) is not observed to low detection limits, which appears to constrain the amount of Si in SiC dust to less that about 5\% \citep{Whittet+Duley+Martin_1990}. However, the SiC absorption profile is highly shape dependent, and irregularly shaped SiC grains could be abundant despite the non-detection at 11.2\,$\mu$m. If the observed ``shoulder'' of the 9.7\,$\mu$m feature is attributed to irregular SiC grains, as much as 9--12\% of the interstellar Si could be in the form of SiC \citep{Whittet+Duley+Martin_1990}.

Little substructure has been detected in the 9.7\,$\mu$m silicate feature, indicating that the feature arises predominantly from amorphous rather than crystalline silicates. Toward Cyg~OB2-12, \citet{Bowey+Adamson+Whittet_1998} found minimal evidence for fine structure between 8.2 and 11.7\,$\mu$m except a possible weak feature at 10.4\,$\mu$m that may be attributable to crystalline serpentine. Measuring silicate absorption toward two protostars and finding a lack of fine structure, \citet{Demyk+etal_1999} determined that at most 1-2\% of the mass of the silicates giving rise to the feature in star-forming clouds could be crystalline, whereas \citet{Kemper+Vriend+Tielens_2005} estimated that at most 2.2\% of the silicate mass in the diffuse ISM could be crystalline. On the basis of detections of the 11.1\,$\mu$m feature from crystalline forsterite in many interstellar environments, \citet{DoDuy+etal_2020} concluded that $\sim1.5$\% of the silicate mass in the diffuse ISM is crystalline, which is consistent with previously derived upper limits. To the extent that the weak, broad 11.1\,$\mu$m feature is present in the extinction toward Cyg\,OB2-12, it is implicitly included in the representative extinction curve we derive in this work.

\subsection{Carbonaceous Features}
\label{subsec:ext_carbon_features}
The presence of extinction features arising from carbon bonds is well-attested in the diffuse ISM. We review here the extinction ``bump'' at 2175\,\AA, the infrared extinction features, and the diffuse interstellar bands (DIBs).

\subsubsection{\texorpdfstring{The 2175\,{\rm \AA}\ Feature}{The 2175A Feature}}
\label{subsubsec:2175}
As evidenced in Figure~\ref{fig:ext_uv}, a striking feature of the interstellar extinction curve is the ``bump'' at 2175\,\AA. This feature was first discovered by \citet{Stecher_1965} and quickly identified with extinction from small graphite particles \citep{Stecher+Donn_1965}, although this identification is not universally accepted. As the backbone of a PAH is in many ways analogous to a graphite sheet, the 2175\,\AA\ feature may be attributable to PAHs \citep{Donn_1968,Draine_1989b,Joblin+Leger+Martin_1992,Draine_2003}.

Regardless of the carrier of the feature, a number of observational facts appear clear. First, the feature appears ubiquitous in the ISM, found over a wide range of $E(B-V)$ \citep{Bless+Savage_1972,Savage_1975}. Second, the feature is quite strong and therefore its carrier must be composed of one of the more abundant elements in the ISM---C, O, Mg, Si, or Fe \citep{Draine_1989b}. Third, the central wavelength of the feature is nearly invariant across many sightlines, though the width can vary dramatically (FWHM between 360 and 600\,\AA) across environments \citep{Fitzpatrick+Massa_1986}. Finally, this feature is weaker, and in some cases absent, in sightlines toward the LMC \citep{Fitzpatrick_1985, Clayton+Martin_1985, Fitzpatrick_1986,Misselt+Clayton+Gordon_1999} and SMC \citep{RoccaVolmerange+etal_1981, Prevot_1984,Thompson+etal_1988, Gordon+etal_2003}.

The consistency of the central wavelength across environments suggests that the feature is relatively insensitive to the grain size distribution, while its weakness in the SMC and LMC lends credence to the idea that it is associated with a specific carrier which may be underabundant in those environments.

While graphite-like sheets, such as those found in PAHs, provide perhaps the most attractive explanation for the feature at present, it is not without difficulties. In particular, \citet{Draine+Malhotra_1993} demonstrated that graphite has difficulty explaining the observed variations in the width of the feature by variations in the size and shape of the grains while simultaneously preserving the constant central wavelength. Alternative hypotheses, such as transitions in OH$^{-}$ ions in amorphous silicates \citep{Steel+Duley_1987}, onion-like carbonaceous composite materials \citep{Wada+etal_1999}, and hydrogenated amorphous carbon \citep{Mennella+etal_1998, Duley+Hu_2012}, provide ways to account for the feature without invoking graphite, though most of these models still attribute the feature to carbonaceous bonds. As of yet, no hypothesis offers a clear explanation for the simultaneous near-invariance of the central wavelength and substantial variation in the feature's width.

\subsubsection{Infrared Features}
\label{subsubsec:carbon_features_mir}

\begin{figure*}
    \centering
        \includegraphics[width=\textwidth]{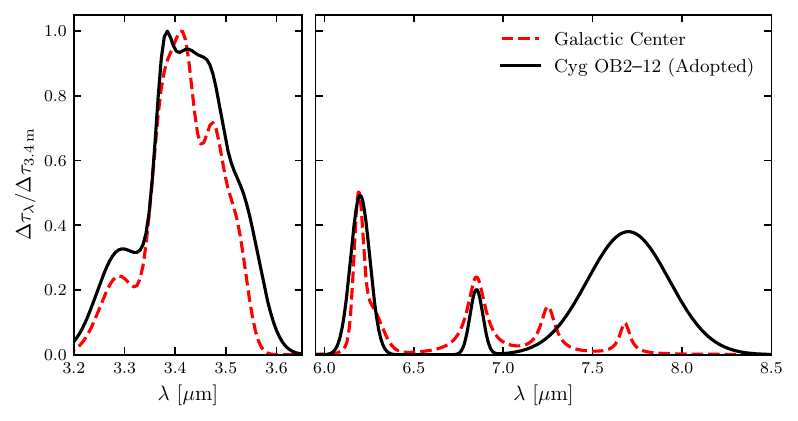}
    \caption{We compare determinations of the hydrocarbon feature profiles toward the Galactic Center based on \replaced{ISO-{\it SWS}}{{\it ISO}-SWS} spectroscopy \citep{Chiar+etal_2000,Chiar+etal_2013} and toward Cyg\,OB2-12 which employed both ISO-{\it SWS} and {\it Spitzer} IRS spectroscopy \citep{Hensley+Draine_2020}. Both sets of profiles have been normalized to the maximum optical depth in the 3.4\,$\mu$m feature. Although these sightlines probe very different interstellar environments, the agreement is excellent aside from the 7.25 and 7.7\,$\mu$m features, where the determinations are most uncertain.} \label{fig:ext_ch}
\end{figure*}

An interstellar absorption feature at 3.4\,$\mu$m was first discovered by \citet{Soifer+Russell+Merrill+1976} toward the Galactic Center source IRS7, though it was not until the non-detection of emission features at 6.2 and 7.7\,$\mu$m along the same line of sight that its interstellar origin was appreciated \citep{Willner+etal_1979}. \citet{Wickramasinghe+Allen_1980} detected a pronounced 3.4\,$\mu$m feature toward IRS7 as well as toward the M star OH\,01--477, which they attributed to the CH stretch band. Detection of this feature toward \replaced{Cyg\,OB-12}{Cyg\,OB2-12} suggests that it is a generic feature of extinction from the diffuse ISM \citep{Adamson+Whittet+Duley_1990,Whittet+etal_1997}.

Subsequent observations of the 3.4\,$\mu$m feature revealed a complex profile, including a number of ``subpeaks'' at 3.39, 3.42, and 3.49\,$\mu$m \citep{Duley+Williams_1983, Butchart+etal_1986, Sandford+etal_1991}. \citet{Sandford+etal_1991} demonstrated consistency between these features and C-H stretching in $=$CH$_2$ (methylene) and -CH$_3$ (methyl) groups in aliphatic hydrocarbons. These results were supported by the more extensive observations of \citet{Pendleton+etal_1994}, who determined that diffuse ISM has a characteristic CH$_2$ to CH$_3$ abundance of about 2.0--2.5. Detailed comparison of the 3.4\,$\mu$m feature to laboratory measurements of a range of materials yielded a close match with hydrocarbons with both aliphatic and aromatic characteristics \citep{Pendleton+Allamandola_2002}.

A key prediction of the aliphatic hydrocarbon origin of the 3.4\,$\mu$m feature is the presence of a 6.85\,$\mu$m CH deformation mode. \citet{Tielens+etal_1996} identified this feature in an IR spectrum of the Galactic Center, confirming this hypothesis. Additionally, they identified features at 5.5 and 5.8\,$\mu$m with C=O (carbonyl) stretching and a feature at 5.5\,$\mu$m with metal carbonyls such as Fe$\left({\rm CO}\right)_4$. Subsequently, \citet{Chiar+etal_2000} detected a 7.25\,$\mu$m feature ascribed to a methylene deformation mode toward the Galactic Center. The 6.85\,$\mu$m feature has been observed toward Cyg\,OB2-12 with the same strength relative to the 3.4\,$\mu$m feature as seen toward the Galactic Center \citep{Hensley+Draine_2020}. Thus, the 6.85\,$\mu$m feature also appears generic to extinction from the diffuse ISM. On the other hand, the 7.25\,$\mu$m feature was {\it not} detected toward Cyg\,OB2-12, although a weak feature could not be completely ruled out. The hydrocarbon feature profiles toward the Galactic Center and Cyg\,OB2-12 are compared in Figure~\ref{fig:ext_ch}.

The 3.47\,$\mu$m subfeature of the 3.4\,$\mu$m complex has been attributed to bonds between H and $sp^3$ bonded (diamond-like) C \citep{Allamandola+etal_1992}. This feature appears to be present in the spectrum of the Galactic Center \citep{Chiar+etal_2013} and absorption in the vicinity of this feature is even stronger toward Cyg\,OB2-12 \citep{Hensley+Draine_2020}, as illustrated in Figure~\ref{fig:ext_ch}. While this suggests diamond-like C may be ubiquitous in both the dense and diffuse ISM, it is in conflict with the finding of \citet{Brooke+etal_1996} that the strength of the 3.47\,$\mu$m feature is better correlated with the 3.1\,$\mu$m H$_2$O ice feature (absent toward Cyg\,OB2-12) than with the 9.7\,$\mu$m silicate feature. Observations of these features on more sightlines are needed to clarify the evolution of hydrocarbons in the ISM.

The attribution of the strong IR emission features to PAHs (see Section~\ref{sec:pah_emission}) implies the presence of aromatic features in the interstellar extinction curve in addition to the observed aliphatic features. Observing eight IR sources, including two Galactic Center sources and Cyg\,OB2-12, with {\it ISO}-SWS spectroscopy, \citet{Schutte+etal_1998} detected a 6.2\,$\mu$m absorption feature associated with aromatic hydrocarbons, which has a well-known corresponding emission feature. Subsequently, both the 3.3 and 6.2\,$\mu$m aromatic features were detected in absorption toward the Quintuplet Cluster \citep{Chiar+etal_2000,Chiar+etal_2013}, and \citet{Hensley+Draine_2020} reported detections of the 3.3, 6.2, and 7.7\,$\mu$m aromatic features in absorption toward Cyg\,OB2-12. While there is a feature in the extinction curve toward the Galactic Center in the vicinity of 7.7\,$\mu$m, \citet{Chiar+etal_2000} attributed it to the 7.68\,$\mu$m feature from methane ice. Because of the detection on the iceless sightline toward Cyg\,OB2-12, we include it in Figure~\ref{fig:ext_ch}, but note that there are substantial observational uncertainties on the depth and width of the feature in both the Galactic Center and Cyg\,OB2-12 determinations \added{that may account for the difference}. \added{However, it is also possible that the discrepancy is real and that there exist fundamental differences in the relative strengths of the extinction features on these two very different sightlines.} The strength of the 7.7\,$\mu$m feature detected toward Cyg\,OB2-12 is, however, consistent with predictions of models for interstellar PAHs \citep{Draine+Li_2007}.

While the aromatic 3.3\,$\mu$m feature is substantially weaker than the aliphatic 3.4\,$\mu$m feature in absorption, it dominates in emission. It is also noteworthy that the 3.3\,$\mu$m feature width is substantially broader in absorption \citep[$\Delta\lambda^{-1} \simeq 90\,$cm$^{-1}$,][]{Chiar+etal_2013,Hensley+Draine_2020} than seen in emission \citep[$\Delta\lambda^{-1} \simeq 30\,$cm$^{-1}$,][]{Tokunaga+etal_1991,Joblin+etal_1996}.


As with the silicate features, carbonaceous features {\it not} observed in the diffuse ISM also constrain dust composition. Polycrystalline graphite is expected to have a lattice resonance in the vicinity of 11.53\,$\mu$m \citep{Draine_1984,Draine_2016}. Such a feature was not observed towards Cyg\,OB2-12 \citep{Hensley+Draine_2020}, though the weakness of the feature allowed only an upper limit of $<$160\,ppm of C in graphite to be set. More stringent upper limits will require more sensitive data and possibly a sightline without contaminating H recombination lines.

Laboratory data suggest the presence of NIR features at 1.05 and 1.23\,$\mu$m associated with ionized PAHs having 40--50 C atoms \citep{Mattioda+etal_2005a,Mattioda+etal_2005b}. These wavelengths may be too short for even ultrasmall grains to produce strong emission features, but if present they should be observable in extinction \citep{Mattioda+etal_2005a}. However, we are unaware of any existing observational constraints on the presence or absence of these features.

\subsubsection{The Diffuse Interstellar Bands}
The diffuse interstellar bands are a set of numerous, relatively broad (hence ``diffuse'') interstellar absorption features that likely arise from molecular transitions. The first two DIBs $\lambda5780$ and $\lambda5795$ were noted as unidentified stellar absorption features \citep{Heger_1922a, Heger_1922b}, but their interstellar nature was not confirmed until \citet{Merrill_1936} found that the lines remained at fixed wavelength in a spectroscopic binary while the stellar lines exhibited the expected time-dependent oscillation. Subsequently, over five hundred DIBs have been identified, the vast majority of which have not been identified with a specific molecular carrier \citep{Herbig_1995, Hobbs+etal_2009, Fan+etal_2019}.

The first definitive identification of a DIB carrier did not occur until 2015 when laboratory measurements demonstrated that C$_{60}^+$ can reproduce the absorption features at 9632 and 9577\,\AA\ \citep{Campbell+etal_2015}. Subsequent detection of the predicted 9428\,\AA\ band has confirmed C$_{60}^+$ as the carrier \citep{Cordiner+etal_2019}. Based on the observed DIB strength, it is estimated that C$_{60}^+$ accounts for only $\sim0.1$\% of the interstellar carbon abundance \citep{Berne+etal_2017}.

The correlation between DIB strength and total reddening is non-linear \citep{Snow+Cohen_1974} and varies among DIBs, suggesting that the various DIB carriers preferentially reside in different interstellar environments, e.g., atomic versus molecular gas \citep{Lan+etal_2015}. It is in principle possible to construct a representative spectrum for DIBs in diffuse \ion{H}{i} gas assuming the empirical relations between DIB equivalent widths and $N_\ion{H}{i}$ derived by \citet{Lan+etal_2015} for the set of 20 DIBs between 4430 and 6614\,\AA\ considered in their study, but we do not pursue such an undertaking in this work.

\subsection{Other Features}
Although we have discussed a number of extinction features associated with specific materials found in diffuse interstellar gas, this inventory is incomplete, particularly as we push to weaker features. Indeed, \citet{Massa+etal_2020} recently presented \added{new} evidence of ``Intermediate Scale Structure,'' i.e., extinction features a few hundred to 1000\,\AA\ wide \added{\citep[see also][]{York_1971}}, in the spectrophotometric extinction curves of \citet{Fitzpatrick+etal_2019}. They identified two features at 4370 and 4870\,\AA\ which both showed correlation with the strength of the 2175\,\AA\ feature, and one feature at 6300\,\AA\ which did not. Further, they argue that the reported ``Very Broad Structure'' \citep{Whiteoak_1966} is actually a minimum between the 4870 and 6300\,\AA\ features. These features affect the optical extinction at the $\lesssim10\%$ level, and we include them in our representative extinction curve only insofar as they are inherent in the mean extinction curves of \citet{Schlafly+etal_2016} and \citet{Fitzpatrick+etal_2019}, which we employ over this wavelength range.

\subsection{\texorpdfstring{$N_{\rm H}/E(B-V)$}{NH/E(B-V)}}
\label{subsec:nh_ebv}
It is expected that the amount of extinction on a given sightline scales linearly with the dust column density and, to the extent that dust and gas are well mixed, with the gas column density. This scaling is borne out observationally and is typically summarized by the quantity $N_{\rm H}/E(B-V)$, which appears roughly constant for the diffuse ISM. Using Ly$\alpha$ absorption measurements made by the {\it Copernicus} satellite for 75 stars within 3400\,pc, \citet{Bohlin+Savage+Drake_1978} derived a value of $N_{\rm H}/E(B-V) = 5.8\times10^{21}$\,H\,cm$^{-2}$\,mag$^{-1}$. They noted that very few of their sightlines differ from this relation by more than a factor of 1.5. Ly$\alpha$ absorption studies with IUE by \citet{Shull+vanSteenberg_1985} and \citet{Diplas+Savage_1994} derived similar $N_\ion{H}{i}/E(B-V)$ values of 5.2 and 4.9$\times10^{21}$\,H\,cm$^{-2}$\,mag$^{-1}$, respectively. Finally, \citet{Rachford+etal_2009} obtained $N_{\rm H}/E(B-V) = \left(5.94\pm0.37\right)\times10^{21}$\,H\,cm$^{-2}$\,mag$^{-1}$ with data from {\it FUSE} for translucent clouds ($A_V \gtrsim 0.5)$ where both \ion{H}{i} and H$_2$ were measured directly.

Measuring the \ion{H}{i} column density toward globular clusters using the 21\,cm line, \citet{Knapp+Kerr+1974} and \citet{Mirabel+Gergely_1979} found $N_\ion{H}{i}/E(B-V)$ of 5.1 and 4.6$\times10^{21}$\,H\,cm$^{-2}$\,mag$^{-1}$, respectively. These values are also consistent with data from a similar study using RR Lyrae \citep{Sturch_1969}, all of which corroborate the values from \ion{H}{i} absorption studies.

However, employing 21\,cm data from the Leiden-Argentina-Bonn (LAB) \ion{H}{i} Survey \citep{Kalberla+etal_2005} and the Galactic Arecibo L-band Feed Array (GALFA) H{\sc i} Survey \citep{Peek+etal_2011} in conjunction with the reddening map of \citet{Schlegel+Finkbeiner+Davis_1998}, \citet{Liszt_2014a} determined $N_\ion{H}{i}/E(B-V) = 8.3\times10^{21}$\,cm$^{-2}$\,mag$^{-1}$ for $|b| \gtrsim 20^\circ$ and $E(B-V) \lesssim 0.1$\, mag. This is a factor of 1.4 higher than that found by \citet{Bohlin+Savage+Drake_1978}. \citet{Liszt_2014a} noted that some previous determinations using \ion{H}{i} emission are in good agreement with this higher value, particularly for $E(B-V) < 0.1$. For instance, \citet{Heiles_1976} found $E(B-V) = (-0.041 \pm 0.012) + N_\ion{H}{i}/(4.85 \pm 0.36) \times10^{21}$\,cm$^{-2}$\,mag$^{-1}$, consistent with the higher value of \citet{Liszt_2014a} when $E(B-V) < 0.1$ due to the negative intercept. Likewise, \citet{Mirabel+Gergely_1979} required a negative intercept to fit their data, suggesting a change in behavior at low reddening.

In a subsequent analysis, \citet{Lenz+Hensley+Dore_2017} correlated $N_\ion{H}{i}$ measurements from the HI4PI Survey \citep{HI4pi_2016} and maps of interstellar reddening as determined by \citet{Schlegel+Finkbeiner+Davis_1998} over the diffuse, high-latitude sky. They found a characteristic $N_\ion{H}{i}/E(B-V) = 8.8\times10^{21}$\,cm$^{-2}$\,mag$^{-1}$ on these sightlines, with a systematic uncertainty of about 10\%. Comparing 21\,cm observations to stellar extinction along 34 sightlines with little molecular gas, \citet{Nguyen+etal_2018} found a compatible $N_{\rm H}/E(B-V) = \left(9.4\pm1.6\right)\times10^{21}$\,cm$^{-2}$\,mag$^{-1}$ (95\% confidence interval) and that this relation persists to $N_{\rm H}$ as high as $3\times10^{21}$\,cm$^{-2}$. Using X-ray absorption to infer $N_{\rm H}$, \citet{Zhu+etal_2017} found a mean value of $N_{\rm H}/A_V = \left(2.08\pm0.02\right)\times10^{21}$ toward a sample of supernova remnants, planetary nebulae, and X-ray binaries across the Galaxy. For $R_V = 3.1$, this corresponds to $N_{\rm H}/E(B-V) = 6.45\times10^{21}$\,cm$^{-2}$\,mag$^{-1}$, intermediate between the \citet{Bohlin+Savage+Drake_1978} and \citet{Lenz+Hensley+Dore_2017} values.

The striking difference between these different determinations of $N_{\rm H}/E(B-V)$ is consistent with systematic variations of the dust-to-gas ratio in the Galaxy, with more dust per H atom in the Galactic plane and less at high Galactic latitudes. As we focus here on high latitude sightlines where the dust emission per H atom is best determined (see Section~\ref{sec:irem}), we adopt the value $N_{\rm H}/E(B-V) = 8.8\times10^{21}$\,cm$^{-2}$\,mag$^{-1}$ of \citet{Lenz+Hensley+Dore_2017} as our benchmark.

\subsection{Scattering}
Extinction is the sum of two processes---absorption and scattering. The scattering properties of dust can be constrained by studying the surface brightness profile of scattered light around point sources and the spectrum of the diffuse Galactic light. However, both of these constraints involve simultaneous modeling of both the dust optical properties and the scattering geometry and are therefore difficult to incorporate self-consistently into the present analysis. We provide a brief overview below, but do not \replaced{incorporate}{include} these observations into our final set of model constraints.

\subsubsection{X-ray Scattering}
\label{subsubsec:xray_sca}

Interstellar grains scatter X-rays through small angles \citep{Overbeck_1965,Hayakawa_1970,Martin_1970}, which can be observed as a ``scattering halo'' in X-ray images of point sources with intervening interstellar dust \citep{Catura_1983,Mauche+Gorenstein_1986}. The scattering is sensitive to both dust composition and size distribution, providing additional observational constraints that a grain model should satisfy.

The angular extent of the scattering halo also depends on the location of the dust between us and the source. For Galactic sources (e.g., low-mass X-ray binaries), this introduces uncertainty when comparing models to observations.

The best-studied case is GX~13+1 \citep{Smith_2008}. \citet{Valencic+Smith_2015} surveyed 35 X-ray scattering halos, and concluded that most could be satisfactorily fit by one or more dust models with size distributions having few grains larger than $\sim0.4\,\mu$m. Extragalactic sources with intervening Galactic dust, the exact distance to which would be unimportant, would be optimal for testing dust models, but high signal-to-noise imaging of X-ray halos around reddened AGN is lacking.

The scattering cross section for the dust grains is expected to show spectral structure near X-ray absorption edges \citep{Draine_2003b}.  If this could be observed, it would provide a means to detect or constrain variations of grain composition with size. \citet{Costantini+etal_2005} reported spectral structure in the scattering halo around Cyg~X-2. Features appear to be present near the O K, Fe L, Mg K, and Si K absorption edges, although the interpretation remains unclear. Future X-ray telescopes may enable more sensitive spectroscopy of scattering halos. 

A population of aligned, aspherical grains can produce observable asymmetries in an X-ray scattering halo \citep{Draine+AllafAkbari_2006}. \citet{Seward+Smith_2013} employed {\it Chandra} observations of Cyg~X-2 to search for these asymmetries, but found the X-ray halo to be uniform in surface brightness to at least the 2\% level. A detection of halo asymmetry has yet to be reported.

Because X-ray scattering is sensitive to grain structure on small scales, X-ray halos can also provide constraints on grain porosity. Analyzing the {\it Chandra} observations of the Galactic binary GX13+1 of \citet{Smith+Edgar+Shafer_2002}, \citet{Heng+Draine_2009} found that the small angle scattering from grains with porosity greater than 0.55 overpredicts the observed surface brightness in the core of the scattering halo. As the degree of compactness of interstellar grains remains a major unresolved question, ancillary data and analysis is needed to test the conclusions of \citet{Heng+Draine_2009}.

\subsubsection{Diffuse Galactic Light}
\begin{figure}
    \centering
        \includegraphics[width=\columnwidth]{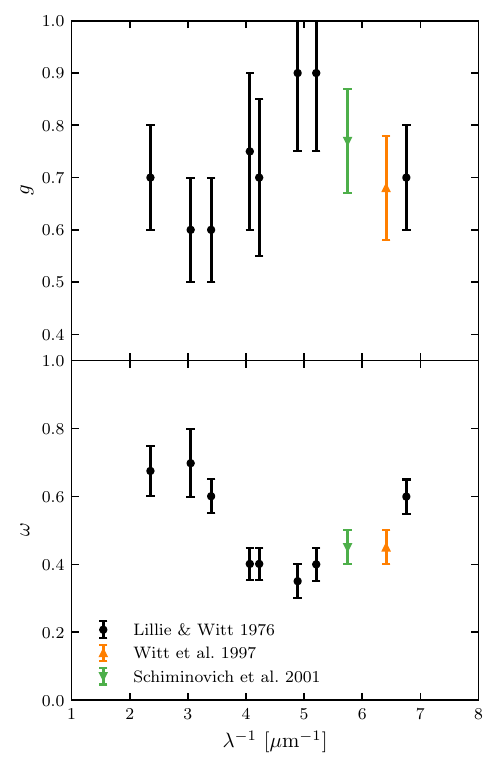}
    \caption{Constraints on the dust albedo $\omega$ and phase function asymmetry $g \equiv \langle \cos\theta \rangle$ inferred from measurements of the diffuse Galactic light.} \label{fig:scattering} 
\end{figure}

Even in a dark patch of sky far from point sources, there is still light from emission from the ISM and from starlight that has been scattered off of dust grains. This ``diffuse Galactic light'' (DGL) was first detected in the photoelectric measurements of \citet{Elvey+Roach_1937}, who derived a surface brightness of 5.6\,mag per square degree at $\lambda \approx 4500$\,\AA. These results were corroborated by the photometric observations of \citet{Henyey+Greenstein_1941}, who concluded that dust grains must have a relatively large albedo $\omega$ ($0.3 < \omega < 0.8$) and be relatively forward scattering, having anisotropy parameter $g \equiv \langle \cos\theta \rangle$ (where $\theta$ is the scattering angle) greater than 0.65. Particles in the Rayleigh limit (i.e., small compared to the wavelength) have $g \approx 0$, i.e., isotropic scattering, indicating that the scattering in the ISM is dominated by larger grains (radius $a \gtrsim 0.1\,\mu$m).

The conversion of measurements of the intensity of scattered light into constraints on the scattering properties of interstellar dust is challenging as it requires assumptions on the distribution of both sources and scatterers. Nevertheless, observations of the DGL from the optical to the UV have been used to constrain the wavelength dependence of both $\omega$ and $g$.

Employing 1500--4200\,\AA\ photometric observations from the Orbiting Astronomical Observatory (OAO-2) in 71 fields at varying Galactic longitude, \citet{Lillie+Witt_1976} found good agreement with earlier ground-based measurements of the DGL. They constrained $\omega$ and $g$ through a radiative transfer analysis on a plane-parallel galaxy in which both dust and stars decrease exponentially with height above the disk, finding $0.3 < \omega < 0.7$ with indications of a minimum near 2200\,\AA, coincident with the extinction bump (see Section~\ref{subsubsec:2175}). Except in this minimum where $g$ attained values as high as 0.9, they found $0.6 < g < 0.7$. 

The UV spectrometers aboard the two {\it Voyager} spacecraft were used to study dust scattering in the Coalsack Nebula by \citet{Murthy+Henry+Holberg_1994}. They employed a simple scattering model assuming fixed $g$ and single scattering only to infer the wavelength dependence of the dust albedo. Fixing $\omega = 0.5$ at 1400\,\AA, they computed the {\it relative} albedo at other wavelengths, finding little wavelength dependence aside from a modest increase toward shorter wavelengths. A follow-up analysis by \citet{Shalima+Murthy_2004} using a more sophisticated Monte Carlo model for the dust scattering determined the FUV dust albedo to be $0.4\pm0.2$.

The Far Ultraviolet Space Telescope (FAUST) measured the diffuse UV continuum between 140 and 180\,nm. Employing the 156\,nm flux measurements from this experiment and a radiative transfer model that accounted for non-isotropic radiation fields and multiple scatterings, \citet{Witt+Friedmann+Sasseen_1997} derived a FUV dust albedo of $0.45\pm0.05$ and $g = 0.68\pm0.10$. The rocket-borne Narrowband Ultraviolet Imaging Experiment for Wide-Field Surveys (NUVIEWS) measured the diffuse UV background at 1740\,\AA. Using a 3D Monte Carlo scattering model based on that described in \citet{Witt+Friedmann+Sasseen_1997}, \citet{Schiminovich+etal_2001} constrained the dust albedo to be $\omega = 0.45\pm0.05$ and $g = 0.77\pm0.1$.

By correlating the spectra of SDSS sky fibers (i.e., spectra of the ``blank'' sky taken for calibration purposes) against the 100\,$\mu$m dust emission measured by IRAS, \citet{Brandt+Draine_2012} measured the spectrum of the DGL between 3900 and 9200\,\AA. Modeling the DGL scattering geometry with a plane-parallel exponential galaxy, they compared the observed spectrum to predictions from dust models. Their formalism could in principle be used to place constraints directly on $\omega$ and $g$, but we do not pursue such analysis here.

We summarize these constraints on the dust albedo and asymmetry parameter in Figure~\ref{fig:scattering}. Given the modeling uncertainties inherent in translating the DGL intensity to the scattering properties of interstellar dust, we do not at this time incorporate these data into our set of constraints. These limitations notwithstanding, it is clear that interstellar dust must have a UV/optical albedo of order 0.5 and be relatively forward scattering ($g > 0.5$).

\subsection{Spatial Variation of the Extinction Curve}
\label{subsec:ext_variations}
It is well established that there is not a single universal extinction curve that describes all regions of the ISM, but rather a variety of extinction curves typically parameterized by $R_V$ \citep{Johnson+Borgman_1963,Cardelli+Clayton+Mathis_1989}. For instance, measurements of extinction toward the Galactic Bulge have indicated $R_V \approx 2.5$ \citep{Udalski_2003,Nataf+etal_2013}. \citet{Schlafly+etal_2016} found large scale gradients in $R_V$, with a follow-up study indicating a possible dependence on Galactocentric radius such that the outer Galaxy has systematically higher $R_V$ than the inner Galaxy \citep{Schlafly+etal_2017}. The magnitude of the variations in  $R_V$, however, was relatively small \citep[$\sigma_{R_V} = 0.18$,][]{Schlafly+etal_2016}. Extinction in dark clouds differs systematically from the diffuse ISM due to the growth of grains by coagulation and the formation of ice mantles. We do not attempt to summarize the observed range of variations in this work, instead restricting our focus to the extinction curve of the local diffuse ISM having an average $R_V \approx 3.1$ \citep{Morgan+etal_1953,Schultz+Wiemer_1975, Sneden+etal_1978, Koornneef_1983, Rieke+Lebofsky_1985,Fitzpatrick+etal_2019}.

\section{Polarized Extinction}
\label{sec:extpol}
Following the discovery that starlight is polarized \citep{Hiltner_1949a, Hiltner_1949b, Hiltner_1949c, Hall_1949, Hall+Miksell_1949, Hall+Mikesell_1950}, it was quickly realized that this polarization was due to selective extinction by aligned dust grains rather than inherent polarization of the stars themselves. \citet{Davis+Greenstein_1951} proposed a physical model of grain alignment whereby aspherical dust grains were aligned by the local magnetic field. Our understanding of the alignment processes of dust grains has since undergone significant evolution \citep[see][for a review]{Andersson+Lazarian+Vaillancourt_2015}, though it remains clear that observations of polarized extinction constrain the size, shape, composition, and alignment properties of interstellar dust.

In this section we summarize observations of the polarized extinction, focusing upon its wavelength dependence, spectral features, and amplitude per unit reddening.

\subsection{Wavelength Dependence}
\label{sec:extpol_wav}

\begin{figure}
    \centering
        \includegraphics[width=\columnwidth]{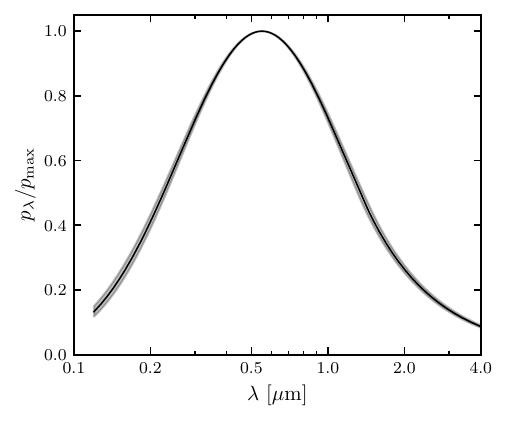}
    \caption{\replaced{We plot the}{The} wavelength dependence of the polarized
      extinction, \replaced{normalizing}{normalized} to the peak polarization. We employ the Serkowski Law in the UV and optical with $\lambda_{\rm max} = 0.55\,\mu$m, and we match this smoothly onto a power law in the IR such that $p_\lambda \propto \lambda^{-1.6}$. The solid line corresponds to a Serkowski Law parameter $K = 0.87$, while the shaded region illustrates the effects of varying $K$ between 0.82 and 0.92, corresponding to the UV- and IR-optimized forms of the Wilking Law described by \citet{Whittet_2003}.} \label{fig:ext_pol} 
\end{figure}

Initial observation of the polarized extinction from UV to NIR wavelengths \citep[e.g.,][]{Behr_1959,Gehrels_1960,Coyne+etal_1974,Gehrels_1974,Serkowski+Mathewson+Ford_1975} established a characteristic wavelength dependence of the polarized extinction that is often parametrized by the ``Serkowski Law'' \citep{Serkowski_1973}:

\begin{equation}
\label{eq:serkowski}
p_\lambda/p_{\rm max} \simeq {\rm
  exp}\,\left[-K\ln^2\left(\lambda_{\rm max}/\lambda\right)\right]
~~~,
\end{equation}
where $p_\lambda$ is the polarization fraction of the two linear polarization modes and $p_{\rm max}$ is the maximum value of $p_\lambda$ occurring at wavelength $\lambda_{\rm max}$. \citet{Serkowski_1971} prescribed the values $K = 1.15$ and $\lambda_{\rm max} = 0.55$\,$\mu$m.

Subsequent observations of polarized extinction revealed that the polarization peak becomes narrower (i.e., $K$ increases) as $\lambda_{\rm max}$ increases \citep{Wilking+etal_1980,Wilking+etal_1982}. This relation, known as the ``Wilking Law,'' is parametrized by the linear relationship

\begin{equation}
K \simeq c_1 \lambda_{\rm max} + c_2
~~~,
\end{equation}
where $c_1$ and $c_2$ are constants to be fit. Analyzing the polarized extinction from the $U$ to $K$ band, \citet{Whittet+etal_1992} derived values of $c_1 = 1.66\,\mu$m$^{-1}$ and $c_2 = 0.01$. Employing UV polarimetry from the Wisconsin Ultraviolet Photo-Polarimeter Experiment (WUPPE), \citet{Martin+Clayton+Wolff_1999} fit values of $c_1 = 2.56\,\mu$m$^{-1}$ and $c_2 = -0.59$. As the former determination is a better fit to the observations from the optical to IR, and the latter a better fit from the UV to the optical, \citet{Whittet_2003} recommended a ``compromise fit'' employing the mean of the two determinations, i.e., $c_1 = 2.11\,\mu$m$^{-1}$ and $c_2 = -0.29$, yielding $K = 0.87$ for $\lambda_{\rm max} = 0.55$\,$\mu$m. For $\lambda_{\rm max} = 0.55\,\mu$m, all three parameterizations produce a similar polarized extinction law, as shown in Figure~\ref{fig:ext_pol}.

Constraints on the polarized extinction law in the UV come almost entirely from WUPPE and the Faint Object Spectrograph on {\it Hubble}, and so while the Serkowski Law appears to describe interstellar polarization down to $\lambda \simeq 1300$\,\AA\ \citep{Somerville+etal_1994}, extrapolations to wavelengths shorter than were accessible by these instruments are uncertain. We therefore adopt 1300\,\AA\ as the shortest wavelength for our polarized extinction curve.

Although the Serkowski Law (Equation~\ref{eq:serkowski}) describes well the polarized extinction in  the UV and optical, it underestimates the observed polarization in the infrared, particularly between $\sim2$ and $5\,\mu$m \citep{Nagata_1990, Jones+Gehrz_1990}. Compiling determinations of the IR polarized extinction along the lines of sight to a number of molecular clouds observed by \citet{Hough+etal_1989}, \citet{Martin+Whittet_1990} determined that the IR polarized extinction could be fit with a power law $p_\lambda \propto \lambda^{-\beta}$ with indices ranging from $\beta = 1.5$ to 2.0. With polarimetry extending from optical wavelengths to 5\,$\mu$m, \citet{Martin+etal_1992} found the $\sim$1--4\,$\mu$m  extinction was well-fit by a power law with index $\beta = 1.6$. Between 4 and 5\,$\mu$m, however, the power law systematically underpredicted the observed polarization.

The behavior of the IR polarized extinction is relatively robust to variations that exist at optical and UV wavelengths as demonstrated by \citet{Clayton+Mathis_1988}. 

For our representative polarized extinction curve, we adopt the ``compromise fit'' of \citet{Whittet_2003} with $K = 0.87$ and $\lambda_{\rm max} = 0.55\,\mu$m from 0.12\,$\mu$m to the infrared. From $\lambda = \lambda_{\rm max} {\rm exp}\left(\beta/2K\right) = 1.38\,\mu$m to 4\,$\mu$m, we adopt a power law with $\beta = 1.6$.

\subsection{Silicate Features}
\label{subsec:sil_features}

\begin{figure}
    \centering
        \includegraphics[width=\columnwidth]{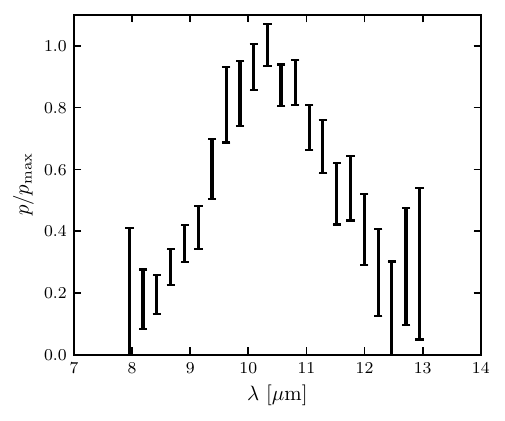}
    \caption{A composite polarized extinction profile of the 9.7\,$\mu$m silicate feature derived by \citet{Wright+etal_2002} from observations toward two Wolf-Rayet stars (WR~48a and WR~112). The extinction on these sightlines appears dominated by the diffuse ISM.} \label{fig:con_silpol} 
\end{figure}

If the features in the interstellar extinction curve arise from aspherical, aligned grains, then these features should also produce polarized extinction. The polarization, or lack thereof, of interstellar extinction features therefore constrains the shape and alignment properties of dust of a specific composition.

The 9.7\,$\mu$m feature was first detected in polarization on the sightline toward the Becklin-Neugebauer (BN) Object in the Orion Molecular Cloud \citep{Dyck+etal_1973, Dyck+Beichman_1974}, with a detection made toward the Galactic Center soon after \citep{Dyck+Capps+Beichman_1974}. Subsequent observations of the BN Object have probed the frequency-dependence of the polarization, including determination of the polarization profile of the 18\,$\mu$m feature \citep{Dyck+Lonsdale_1981, Aitken+etal_1985, Aitken+Smith+Roche_1989}. 

Although the BN Object is well-studied, its molecular environment does not likely typify the diffuse ISM. \citet{Smith+etal_2000} presented an atlas of spectropolarimetry for 55 sources between 8 and 13\,$\mu$m, and, for six of these, additional spectropolarimetric observations between 16 and 22\,$\mu$m. Drawing on these data, \citet{Wright+etal_2002} constructed a typical polarization profile of the 9.7\,$\mu$m silicate feature in extinction based on observations of the Wolf-Rayet stars WR~48a and WR~112. These sightlines were selected because the polarization appears dominated by interstellar absorption. However, both sightlines have H$_2$O ice features at both 3.1 and 6.0\,$\mu$m \citep{Marchenko+Moffat_2017} and so may differ in detail from purely diffuse  sightlines. We present this composite polarization profile in Figure~\ref{fig:con_silpol}.

We are unaware of any published observations that might typify the diffuse ISM along which both 10\,$\mu$m and optical polarimetry have been obtained. Thus, we are unable to normalize the \citet{Wright+etal_2002} polarization profile relative to our polarized extinction curve discussed in Section~\ref{sec:extpol_wav}.

\subsection{Carbonaceous Features}
\label{sec:carbon_extpol}
Unlike the silicate features, the extinction features associated with carbonaceous grains have, with few exceptions, {\it not} been detected in polarization.

The 3.4\,$\mu$m feature is the strongest of the infrared extinction features associated with carbonaceous grains (see Section~\ref{subsubsec:carbon_features_mir}), and as such it is a natural observational target for assessing whether carbonaceous grains give rise to polarized extinction. Low-resolution spectropolarimetric observations of five Galactic Center sources by \citet{Nagata+Kobayashi+Sato_1994} yielded no discernible polarization feature near 3.4\,$\mu$m, nor did high-resolution spectropolarimetric observations of GC-IRS7 by \citet{Adamson+etal_1999}. A subsequent search for the 3.4\,$\mu$m feature in polarization toward the young stellar object IRAS 18511+0146 likewise provided only upper limits \citep{Ishii+etal_2002}.

However, the 9.7\,$\mu$m silicate feature had not been measured along any of these sightlines, leading to ambiguity as to whether the lack of polarization was due to the carbonaceous grains themselves or the magnetic field geometry along the line of sight. This ambiguity was settled by \citet{Chiar+etal_2006} who performed spectropolarimetric observations along two lines of sight in the Quintuplet Cluster which had existing polarimetric measurements of the silicate feature. Finding no evidence of polarization in the 3.4\,$\mu$m feature, they concluded that the carbonaceous grains responsible for the feature are much less efficient polarizers than the silicate grains. Subsequent spectropolarimetric observations of the Seyfert\,2 galaxy NGC\,1068 yielded no detectable feature at 3.4\,$\mu$m \citep{Mason+etal_2007}, supporting the conclusions of \citet{Chiar+etal_2006} in a markedly different interstellar environment and further challenging dust models invoking grains with silicate cores with carbonaceous mantles
\citep[see discussion in][]{Li+Liang+Li_2014}. On the basis of the non-detections reported by \citet{Chiar+etal_2006}, it appears that $\Delta p_{3.4}/\Delta p_{9.7} < 0.03$.

The 2175\,\AA\ feature is a second natural candidate to examine for dichroic extinction arising from carbonaceous grains. Initial WUPPE results suggested excess polarization between 2000 and 3000\,\AA\ on several sightlines, with more detailed modeling suggesting that the excesses toward \replaced{HD\,197770 and HD\,147933-4 ($\rho$ Oph A and B)}{HD\,147933-4 ($\rho$ Oph A and B) and HD\,197770} did in fact arise from the 2175\,\AA\ feature \citep{Clayton+etal_1992,Wolff+etal_1997}. However, if the 2175\,\AA\ feature had the same strength relative to the continuum polarized extinction along all lines of sight, then other detections should have been made, e.g., toward HD\,161056. The sightlines toward \replaced{HD\,197770 and HD\,147933-4}{HD\,147933-4 and HD\,197770} do not betray any unusual behavior in other respects (e.g., the wavelength dependence of the polarization, the extinction curve, etc.), leading \citet{Wolff+etal_1997} to conclude that there are sightline-to-sightline variations in the polarizing efficiency of the grains responsible for the 2175\,\AA\ feature.

It is difficult to draw definitive conclusions on the basis of two detections (and $\sim30$ non-detections), emphasizing the need for more observations of UV polarization on more sightlines. Particularly now that synergy is possible with observations of FIR polarized emission, this effort promises to enhance our understanding of both grain composition and alignment.

\subsection{Maximum \texorpdfstring{$p_V/E(B-V)$}{pV/E(B-V)}}
\label{subsec:pv_ebv}
Interstellar dust grains rotate rapidly with angular momentum preferentially parallel to the local magnetic field. The short axis of each grain tends to align with the angular momentum, and hence is preferentially parallel to the magnetic field. When the line of sight is parallel to the magnetic field, grain rotation eliminates any net polarization. In contrast, the polarization is greatest when the magnetic field is in the plane of the sky. Dust models should reproduce the intrinsic polarizing efficiency of dust grains, and so we focus here on the case of maximal polarization. For dust extinction, this has typically been quantified as the maximum $V$-band polarization per unit reddening, i.e., $\left[p_V/E(B-V)\right]_{\rm max}$.

\citet{Serkowski+Mathewson+Ford_1975} used a sample of 364 stars of various spectral types to derive $\left[p_V/E(B-V)\right]_{\rm max} = 9\%$\,mag$^{-1}$. While individual stars and regions were occasionally found to have $p_V/E(B-V)$ exceeding this upper envelope \citep[e.g.,][]{Whittet+etal_1994, Skalidis+etal_2018}, it was  ambiguous whether dust on these sightlines was atypical or whether the upper envelope had been underestimated. With full-sky polarimetric measurements of dust emission, the {\it Planck} satellite facilitated a detailed comparison between polarized emission in the FIR and polarized extinction in the optical, finding a remarkably linear relation between the submillimeter polarization fraction $p_S$ and $p_V/E(B-V)$ \citep[][see Section~\ref{sec:pol_opt_ir}]{Planck_Int_XXI,Planck_2018_XII}. Given this relationship, the observed $p_S \gtrsim 20\%$ in some regions implies $p_V/E(B-V) \simeq 13\%$\,mag$^{-1}$, leading \citet{Planck_2018_XII} to conclude that the classic envelope of 9\%\,mag$^{-1}$ should be revised.

\citet{Panopoulou+etal_2019} employed $R$-band RoboPol observations of 22 stars in a region with $p_S \gtrsim 20\%$ to find that, indeed, the starlight was polarized in excess of $p_V/E(B-V) = 9\%$\,mag$^{-1}$, perhaps even exceeding 13\%\,mag$^{-1}$. Further, UBVRI polarimetry of six of the 22 stars indicated a typical Serkowski Law in this region, suggesting that the dust on these sightlines is not atypical.

Given these recent observational results, we require that dust models reproduce $p_V/E(B-V) = 13\%$\,mag$^{-1}$, and we normalize our polarization profile to this value.

\section{Emission}
\label{sec:irem}
In this section we review observations of emission from interstellar dust from the infrared to microwave, focusing in particular on the emission per unit H column density characteristic of typical diffuse sightlines.

\subsection{IR Emission}
\label{subsec:irem}

\begin{deluxetable}{ccc}
  \tablewidth{0pc}
      \tablecaption{Infrared Dust Emission Per H\label{table:ir_sed}}
    \tablehead{$\nu$ & $\lambda I_\lambda/N_{\rm H}$
      & $\left(\lambda P_\lambda/N_{\rm H}\right)_{\rm max}$ \\
    $\left[{\rm GHz}\right]$ & $\left[{\rm erg}\,{\rm s}^{-1}\,{\rm
        sr}^{-1}\,{\rm H}^{-1}\right]$ & $\left[{\rm erg}\,{\rm s}^{-1}\,{\rm
        sr}^{-1}\,{\rm H}^{-1}\right]$}
    \startdata
    3000 & $\left(2.05\pm0.29\right)\times10^{-25}$ & \\
    2140 & $\left(2.51\pm0.30\right)\times10^{-25}$ & \\
    1250 & $\left(1.05\pm0.12\right)\times10^{-25}$ & \\
    857 & $\left(3.49\pm0.36\right)\times10^{-26}$ & \\
    545 & $\left(6.78\pm0.73\right)\times10^{-27}$ & \\
    353 & $\left(1.281\pm0.015\right)\times10^{-27}$ &
    $\left(2.514\pm0.030\right)\times10^{-28}$ \\
    217 & $\left(1.698\pm0.016\right)\times10^{-28}$ &
    $\left(3.407\pm0.054\right)\times10^{-29}$ \\
    143 & $\left(2.798\pm0.038\right)\times10^{-29}$ &
    $\left(5.68\pm0.10\right)\times10^{-30}$ \\
    100 & $\left(6.18\pm0.13\right)\times10^{-30}$ &
    $\left(1.174\pm0.031\right)\times10^{-30}$ \\
    94 & $\left(4.66\pm0.22\right)\times10^{-30}$ & \\
    70.4 & $\left(1.544\pm0.050\right)\times10^{-30}$ &
    $\left(2.42\pm0.19\right)\times10^{-31}$ \\
    61 & $\left(9.25\pm0.62\right)\times10^{-31}$ & \\
    44.1 & $\left(5.08\pm0.27\right)\times10^{-31}$ &
    $\left(4.06\pm0.74\right)\times10^{-32}$ \\
    41 & $\left(4.66\pm0.24\right)\times10^{-31}$ & \\
    33 & $\left(4.41\pm0.17\right)\times10^{-31}$ & \\
    28.4 & $\left(4.25\pm0.15\right)\times10^{-31}$ & \\
    23 & $\left(4.05\pm0.13\right)\times10^{-31}$ &   
    \enddata
    \tablecomments{Adopted dust SED per H and maximum polarized SED per H for the high latitude diffuse ISM. These SEDs are based on those presented in \citet{Planck_Int_XVII}, \citet{Planck_Int_XXII}, and \citet{Planck_2018_XI} and have been color corrected (see Section~\ref{subsec:irem}).}
\end{deluxetable}

\begin{figure*}
    \centering
        \includegraphics[width=\textwidth]{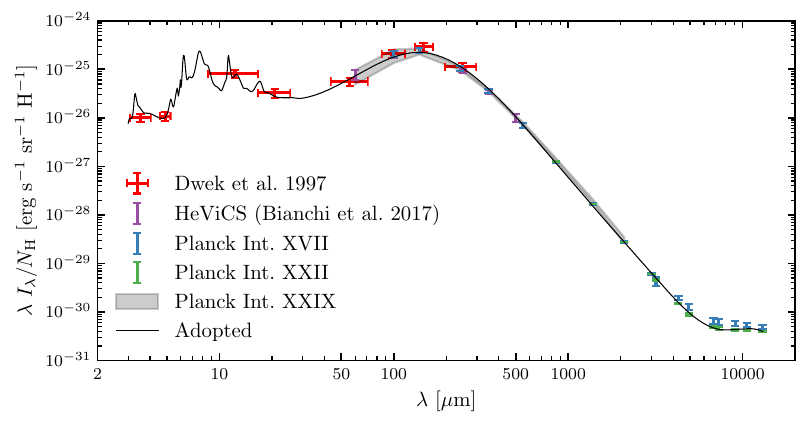}
    \caption{We plot \replaced{two}{several} determinations of the \ion{H}{i}-correlated dust emission from near-IR wavelengths through the microwave. In red is the SED of \citet{Dwek+etal_1997} derived from DIRBE data, \added{in purple the determination of \citet{Bianchi+etal_2017} using {\it Herschel} and IRAS data}, and in blue the SED of \citet{Planck_Int_XVII} which employs DIRBE, WMAP, and {\it Planck} data. The horizontal \replaced{errors}{bars} on the DIRBE data indicate the bandpasses. We plot in gray a range of dust SEDs per $A_V$ from \citet{Planck_Int_XXIX} which we have renormalized to the hydrogen column (see text), and in green a dust SED based on correlations with the 353\,GHz emission \citep{Planck_Int_XXII}. The anomalous microwave emission (AME) is evidenced by the flattening of the SED at wavelengths $\lambda \gtrsim 6\,$mm.} \label{fig:ir_obs}
\end{figure*}

In radiation fields typical of the diffuse ISM, the bulk of the dust grains are heated to $\sim$20\,K and therefore emit thermally in the far-infrared. These wavelengths are largely inaccessible from the ground, necessitating balloon- and space-based observations.

The DIRBE and FIRAS instruments aboard the Cosmic Background Explorer (COBE) constrained the spectrum of the diffuse ISM from 3.5 to 1000\,$\mu$m. In addition to confirming the presence of PAH emission near 3.5 and 4.9\,$\mu$m, \citet{Dwek+etal_1997} derived the \ion{H}{i}-correlated SED of dust in the diffuse ISM. We plot this SED in Figure~\ref{fig:ir_obs}. We note that these data were color corrected assuming a source spectrum with constant $\lambda I_\lambda$ across the band.

Prior to the release of the {\it Planck} dust maps, several studies synthesized the existing data from COBE and WMAP to produce self-consistent dust SEDs. \citet{Paradis+etal_2011} extracted an area of the sky with $|b| > 6^\circ$ and a FIRAS 240\,$\mu$m intensity greater than 18\,MJy\,sr$^{-1}$, corresponding to a sky fraction of 13.7\%. \citet{Compiegne+etal_2011}, also seeking a composite dust SED in which the emission in each band was determined over the same region of the sky, combined DIRBE, FIRAS, and WMAP observations at high Galactic latitudes ($|b| > 15^\circ$) and low \ion{H}{i} column densities ($N_\ion{H}{i} < 5.5\times10^{20}\,{\rm cm}^{-2}$). The differences between these SEDs and that of \citet{Dwek+etal_1997} are minor at their overlapping wavelengths.

The {\it Planck} satellite made sensitive measurements of the FIR-submillimeter dust emission over the full sky. Combining the {\it Planck} data with WMAP and DIRBE and correlating with \ion{H}{i} emission measured by the Parkes 21\,cm survey, \citet{Planck_Int_XVII} constructed a mean SED of the diffuse ISM ($N_\ion{H}{i} \sim 3\times10^{20}$\,cm$^{-2}$) from infrared to microwave wavelengths, which we plot in Figure~\ref{fig:ir_obs}. Following \citet{Planck_Int_XXII}, we apply a correction of 1.9, -2.2, and -3.5\% to the 353, 545, and 857\,GHz bands, respectively, due to updates in the {\it Planck} bandpass determinations subsequent to the work of \citet{Planck_Int_XVII}, and an additional 1.5\% upward correction to the 353\,GHz band following \citet{Planck_2018_XI}. We color correct these data using the tables in \citet{Planck_Int_XVII} to express the SED in terms of monochromatic intensities at the reference frequencies and thus facilitate direct comparison to models. 

Recently, \citet{Planck_Int_LVII} correlated the {\it Planck} 545\,GHz dust amplitude maps from the NPIPE data processing pipeline with \ion{H}{i}4PI maps \citep{HI4pi_2016} filtered to retain only \ion{H}{i} velocities between $\pm90$\,km\,s$^{-1}$ \citep{Lenz+Hensley+Dore_2017}. They found $\lambda I_\lambda/N_{\rm H} = 7.74\times10^{-27}$ erg\,s$^{-1}$\,sr$^{-1}$\,H$^{-1}$ at 545\,GHz, slightly higher than but consistent with the value from \citet{Planck_Int_XVII} quoted in Table~\ref{table:ir_sed}.

The use of \ion{H}{i} correlation to separate the Galactic dust emission from other components becomes increasingly unreliable at low frequencies where these other components, such as free-free and synchrotron, can have non-zero correlation with \ion{H}{i}. \citet{Planck_Int_XXII} derived a microwave dust SED by correlating emission in the lower frequency {\it Planck} bands with the 353\,GHz emission. We plot this SED in Figure~\ref{fig:ir_obs}. While this SED and the \ion{H}{i}-based SED of \citet{Planck_Int_XVII} agree very well from 353 to 94\,GHz, they diverge at lower frequencies.

There is evidence that the shape of the dust SED is not uniform across the sky and indeed varies systematically with the strength of the radiation field that heats the dust. \citet{Planck_Int_XXIX} explored this relationship by fitting the dust model of \citet{Draine+Li_2007} to full-sky maps of infrared dust emission. They then normalized these SEDs to the observed optical extinction based on SDSS observations of more than 250,000 quasars. 

At 353\,GHz, the median SED has an intensity per $A_V$ of 0.92\,MJy\,sr$^{-1}$\,mag$^{-1}$, while \citet{Planck_Int_XVII} measured a 353\,GHz intensity per hydrogen of $3.9\times10^{-22}$\,MJy\,sr$^{-1}$\,cm$^2$\,H$^{-1}$. Taking these at face value implies $A_V/N_{\rm H} = 4.2\times10^{-22}$\,mag\,cm$^2$. In contrast, from our adopted $N_{\rm H}/E(B-V) = 8.8\times10^{21}$\,cm$^{-2}$\,mag$^{-1}$ (see Section~\ref{subsec:nh_ebv}) and $R_V = 3.1$ (see Section~\ref{subsec:ext_op}), we compute $A_V/N_{\rm H} = 3.5\times10^{-22}$\,cm$^2$\,mag. \citet{Green+etal_2018} found that the \citet{Planck_2013_XI} reddening map calibrated on SDSS quasars overpredicted stellar reddenings by a factor of $\sim1.25$ at intermediate latitudes, suggesting these discrepancies are rooted in the reddening calibration. We therefore correct the SEDs per $A_V$ of \citet{Planck_Int_XXIX} upward by 25\% when comparing them to other determinations.

In Figure~\ref{fig:ir_obs}, we plot the range of dust SEDs over different values of the radiation field strength from \citet{Planck_Int_XXIX}. While there are expected systematic variations in the individual SEDs, the range is consistent with the other determinations within the uncertainties. The systematic variations of the dust SED with the radiation field may encode information about the evolution of dust properties in different environments \citep{Fanciullo+etal_2015}.

\added{A final recent determination of the FIR dust emission comes from the {\it Herschel} satellite. \citet{Bianchi+etal_2017} correlated data from the {\it Herschel} Virgo Cluster Survey (HeViCS) with \ion{H}{i} emission data from the Arecibo Legacy Fast ALFA (ALFALFA) Survey over a small patch of sky (76\,deg$^2$) over the Virgo cluster. Despite analyzing only 0.2\% of the sky, they find excellent agreement with the \citet{Planck_Int_XVII} spectrum derived over 7500\,deg$^2$. We include the HeViCS spectrum in Figure~\ref{fig:ir_obs}}.

We adopt as a dust model constraint the dust SED of \citet{Planck_Int_XVII} based on \ion{H}{i} correlation from the 100, 140, and 240\,$\mu$m DIRBE bands, which overlap with the SED of \citet{Dwek+etal_1997}, down to the 353\,GHz {\it Planck} band. Given the known issues with \ion{H}{i} correlation at low frequencies, we adopt the SED of \citet{Planck_Int_XXII} from the {\it Planck} 217\,GHz band to the WMAP 23\,GHz band, normalizing to the measured 353\,GHz intensity per H atom derived by \citet{Planck_Int_XVII}. At the lowest frequencies, the dust emission is dominated by the anomalous microwave emission (AME), which we discuss in Section~\ref{subsec:ame}. Our adopted dust SED is presented in Table~\ref{table:ir_sed}, where we have color corrected all data to facilitate direct comparisons to models.

\subsection{Infrared Emission Features}
\label{sec:pah_emission}
\begin{figure}
    \centering
       \includegraphics[width=\columnwidth]{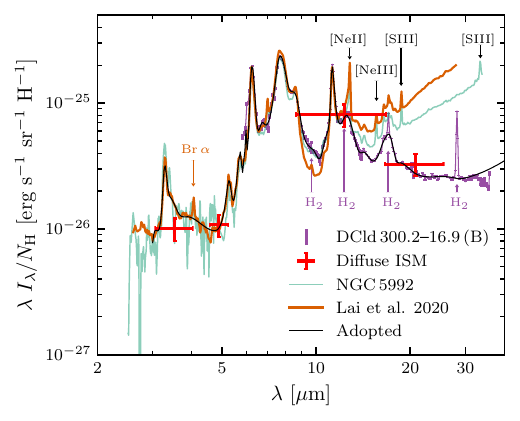}
    \caption{In violet we plot the {\it Spitzer} IRS spectrum of the translucent cloud DCld~300.2-16.9 (B) as determined by \citet{Ingalls+etal_2011}, where we have noted the locations of rotational H$_2$ lines. In \replaced{black}{green} we plot the combined {\it Spitzer} and \replaced{{\it Akari}}{{\it AKARI}} spectrum of the star-forming SBb galaxy NGC\,5992 \citep{Brown+etal_2014}, \added{and in orange we plot the average {\it Spitzer} and {\it AKARI} spectrum of PAH-bright galaxies \citep{Lai+etal_2020}. Both have}\deleted{which has} been corrected for starlight emission by subtraction of a 5000\,K blackbody. We \deleted{also} indicate the strong emission lines present in the \replaced{spectrum}{spectra}. In red, we plot the \ion{H}{i}-correlated dust emission as seen by DIRBE \citep{Dwek+etal_1997}, which we use to normalize the PAH emission spectra.} \label{fig:pah_spectrum} 
\end{figure}

The mid-IR emission from dust is characterized by prominent emission features at 3.3, 6.2, 7.7, 8.6, 11.3, 12.0, 12.7, \replaced{and 13.55\,$\mu$m}{13.55, and 17\,$\mu$m} (see Figures~\ref{fig:ir_obs} and~\ref{fig:pah_spectrum}). First observed in the 1970s \citep[e.g.,][]{Gillett+Forrest+Merrill_1973, Merrill+Soifer+Russell_1975}, these features were subsequently identified as vibrational modes of PAHs \citep{Leger+Puget_1984, Allamandola+Tielens+Barker_1985}. As grains must be heated to quite high temperatures in order to excite these modes ($T \gtrsim 250$\,K), the carriers must be small enough to be heated through the absorption of a single photon. This process can bring small grains to temperatures in excess of 1000\,K.

The width and ubiquity of these emission features make it implausible that they are due to a single species of PAH. Rather, they represent the aggregate emission from a diverse population of PAH-like molecules. The 3.3\,$\mu$m feature, also observed in extinction (see Section~\ref{subsubsec:carbon_features_mir}), has been identified with the aromatic C-H stretching mode; C-C stretching modes account for the 6.2 and 7.7\,$\mu$m features, which have also been observed in extinction (see Section~\ref{subsubsec:carbon_features_mir}); the C-H in-plane bending mode gives rise to the 8.6\,$\mu$m feature, while the C-H out-of-plane bending mode produces the 11.3, 12.0, 12.7, and 13.5\,$\mu$m features depending on whether one, two, three, or four hydrogen atoms are adjacent, respectively. A detailed summary of the features and their corresponding modes can be found in \citet{Allamandola+Tielens+Barker_1989} and \citet{Tielens_2008}.

The strength of these features suggests that a substantial amount of interstellar carbon must reside in the grains giving rise to this emission. The dust model of \citet{Draine+Li_2007} required 4.7\% of the total dust mass to reside in PAHs with fewer than $10^3$ carbon atoms, which accounted for $\sim10$\% of the total interstellar carbon abundance.

In addition to aromatic features associated with PAHs, the aliphatic 3.4\,$\mu$m feature has also been observed in emission \citep[e.g.,][]{Geballe+etal_1985, Sloan+etal_1997,Lai+etal_2020}, though it is typically much weaker than the 3.3\,$\mu$m aromatic feature. Comparing the strengths of these two features and assuming the 3.4\,$\mu$m feature arises solely from aliphatic carbon, \citet{Li+Draine_2012} concluded that no more than about 10\% of the carbon in grains giving rise to these emission features can be in an aliphatic bond. However, it should be noted that anharmonicity in the aromatic 3.3\,$\mu$m C-H stretching mode may also contribute to the emission at 3.4\,$\mu$m \citep{Barker+Allamandola+Tielens_1987, Li+Draine_2012}, further reducing the abundance of the aliphatic component.

Using {\it Spitzer} IRS, \citet{Ingalls+etal_2011} made spectroscopic measurements between 5.2 and 38\,$\mu$m of several regions in the translucent cloud DCld\,300.2-16.9. In addition to \replaced{detecting}{featuring} IR H$_2$ transitions, these measurements provide a reasonable proxy for the PAH emission in the diffuse ISM. We plot the spectrum of their sightline ``B'' in Figure~\ref{fig:pah_spectrum}, where we have noted the observed H$_2$ lines. If a column density of $3.9\times10^{21}\,$cm$^{-2}$ is assumed, the bandpass-integrated SED agrees well with the \ion{H}{i}-correlated DIRBE SED of the diffuse ISM as determined by \citet{Dwek+etal_1997} (see Section~\ref{subsec:irem}). $^{12}$CO observations of this cloud suggest $N({\rm H}_2) \sim 2\times10^{21}$\,cm$^{-2}$ \citep{Ingalls+etal_2011}, and so this column density appears reasonable.

Combining spectroscopy from {\it Spitzer} and \replaced{{\it Akari}}{{\it AKARI}}, along with ancillary data from the UV to the IR, \citet{Brown+etal_2014} presented an atlas of 129 galaxy SEDs spanning a range of galaxy types. We focus on their 2.5--34\,$\mu$m spectrum of NGC\,5992, a star-forming SBb galaxy. To remove the continuum emission from starlight in this spectrum, we subtract a 5000\,K blackbody component. We also note the presence of some emission lines in the spectrum arising from \ion{H}{ii} regions: [\ion{Ne}{ii}] at 12.81\,$\mu$m and [\ion{S}{iii}] at 12.81 and 18.71\,$\mu$m. In Figure~\ref{fig:pah_spectrum}, we compare the MIR spectra of DCld\,300.2-16.9 (B) and NGC\,5992, finding excellent agreement between $\simeq 5$--12\,$\mu$m.

\added{More recently, \citet{Lai+etal_2020} presented a synthesis of combined {\it AKARI} and {\it Spitzer} spectra of star-forming galaxies. In Figure~\ref{fig:pah_spectrum}, we plot their ``1C'' spectrum, corresponding to galaxies that are PAH-bright but excluding active galactic nuclei. As with the NGC\,5992 spectrum, we subtract a 5000\,K blackbody component such that the shape of the spectrum is in rough agreement with the DIRBE SED. This spectrum is in general agreement with both NGC\,5992 and DCld\,300.2-16.9 (B) in regions of overlap, but averaging over many galaxies and an improved treatment of the region of spectral overlap between {\it AKARI} and {\it Spitzer} reveal several features not readily apparent in the NGC\,5992 spectrum, in particular the 5.3\,$\mu$m PAH complex and the subdominant but distinct feature at 3.4\,$\mu$m.}

As the \replaced{AKARI}{{\it AKARI}} data constrain\deleted{s} the PAH emission \deleted{in NGC\,5992} at short wavelengths, we adopt \replaced{this SED}{the SED of \citet{Lai+etal_2020}} as our benchmark between 3 and 12\,$\mu$m. Given the uncertainty of the starlight subtraction, we do not employ the data at wavelengths less than 3\,$\mu$m. The spectra of \replaced{NGC\,5992}{\citet{Lai+etal_2020}} and DCld\,300.2-16.9 diverge beyond 12\,$\mu$m likely due to the more intense starlight heating, and consequently higher temperature grains, in \replaced{NGC\,5992}{the sample of PAH-bright galaxies}. The spectrum of DCld\,300.2-16.9 is more likely to typify the diffuse ISM and is in good agreement with the shape of the DIRBE SED, and thus we adopt it as our benchmark from 12--38\,$\mu$m. However, we excise portions of the spectrum in the vicinity of the S(0), S(1), and S(2) H$_2$ rotational transitions at 28.2, 17.0, and 12.3\,$\mu$m, respectively.

In addition to the hydrocarbon features discussed above, weak mid-infrared emission features from the C-D stretching modes of deuterated aromatic and aliphatic hydrocarbons are expected near 4.5\,$\mu$m, given that in the diffuse ISM D is often substantially depleted from the gas phase \citep{Linsky+etal_2006}. Detections of such emission features have been reported \citep{Peeters+etal_2004,Doney+etal_2016}, but interpretation remains uncertain.

\subsection{Anomalous Microwave Emission}
\label{subsec:ame}
The anomalous microwave emission (AME) was discovered as a dust-correlated emission component in the microwave, both in {\it COBE} maps at 31.5, 53, and 90\,GHz \citep{Kogut+etal_1996,deOliveiraCosta+etal_1997} and observations of the North Celestial Pole made with the Owens Valley Radio Observatory 5.5\,m telescope at 14.5 and 32\,GHz \citep{Leitch+etal_1997}. While these studies suggested free-free emission as a possible explanation, \citet{Draine+Lazarian_1998a} argued against this interpretation on energetic grounds and suggested instead that electric dipole emission from spinning ultra-small grains was the responsible mechanism. For a recent review of AME, see \citet{Dickinson+etal_2018}.

The Perseus Molecular Cloud is perhaps the best-studied AME source and the excellent frequency coverage near the AME peak helps constrain the underlying SED. It exhibits a pronounced emission peak near 30\,GHz with a sharp decline to both higher and lower frequencies \citep[see][for a compilation of low-frequency observations of Perseus]{GenovaSantos+etal_2015}.

The AME of the diffuse ISM appears systematically different than what has been observed in specific clouds. For instance, the AME SED derived from all-sky WMAP and {\it Planck} maps does {\it not} exhibit a low-frequency turnover but rather has a spectrum that appears to rise through the lowest frequency band \citep[WMAP 23\,GHz;][]{MivilleDeschenes+etal_2008, Planck_2015_X}. However, C-BASS observations in the North Celestial Pole region indicate no presence of diffuse AME at 5\,GHz \citep{Dickinson+etal_2019}. More data between 5 and 23\,GHz is required to place constraints on the AME SED of the diffuse ISM, in particular its peak frequency.

The SED of dust-correlated emission derived by \citet{Planck_Int_XXII} and presented in Table~\ref{table:ir_sed} includes an AME component at microwave frequencies, as can be seen in Figure~\ref{fig:ir_obs}. However, the 353\,GHz emission is not perfectly correlated with AME in general \citep[e.g.][]{Planck_Int_XV, Hensley+Draine+Meisner_2016, Planck_2015_XXV, Dickinson+etal_2019}, and so a correlation analysis may underestimate the amount of AME relative to the submillimeter dust emission. Additionally, the other low-frequency foregrounds like free-free and synchrotron emission are also dust-correlated
\citep{Choi+Page_2015, Krachmalnicoff+etal_2018}, which may bias the shape of the derived AME SED.

Parametric component separation with the \texttt{Commander} code has yielded full-sky maps of AME \citep{Planck_2015_X} and mitigates some of the concerns with a correlation-based approach. Employing these maps over the full sky, \citet{Planck_2015_XXV} found the ratio of specific intensities $I_\nu$ of the 22.8\,GHz AME to the 100\,$\mu$m and 545\,GHz dust  emission to be $\left(3.5\pm0.3\right)\times10^{-4}$ and $\left(1.0\pm0.1\right)\times10^{-3}$, respectively. When instead restricting to $|b| > 10^\circ$, consistent results are obtained to within the uncertainties. This agrees reasonably well with the \citet{Planck_Int_XXII} SED, which has corresponding ratios of $2.6\times10^{-4}$ and $1.1\times10^{-3}$, respectively.

Given this agreement, we take the SED of \citet{Planck_Int_XXII} as representative even at AME-dominated wavelengths. However, we note that the AME varies both in intrinsic strength and peak frequency from region to region \citep{Planck_Int_XV, Planck_2015_XXV}, so comparisons between dust models and an average SED should be made with care.

\subsection{Luminescence}

In addition to scattering optical light, dust grains  also luminesce---emit optical photons following absorption of a higher energy photon. This can be the result of fluorescence---radiative deexcitation of the excited electronic level produced by absorption. Alternatively, internal conversion may lead to excitation of a different electronically-excited state that then deexcites radiatively, a process termed ``Poincar\'e fluorescence'' \citep{Leger+etal_1988}.

Luminescence at extreme red wavelengths (6000--8000\,\AA, corresponding to $1.5\lesssim h\nu\lesssim 2.1$\,eV) has been observed in a number of reflection nebulae, including the well-studied objects NGC\,2023 and NGC\,7023 \citep{Witt+Boroson_1990}. Because the emission is spatially extended, it is referred to as ``extended red emission'' \citep[ERE;][]{Witt+Schild+Kraiman_1984}. ERE is also seen in some planetary nebulae \citep{Furton+Witt_1990, Furton+Witt_1992}, and in some unusual systems such as the Red Rectangle  \citep{Cohen+etal_1975,Schmidt+Cohen+Margon_1980}, where it was first discovered. The dust in reflection nebulae is presumed to be interstellar dust that happens to be illuminated by a nearby star, and so we expect ERE to be a property of the general interstellar dust population.

ERE is present in carbon-rich planetary nebulae, but has not been observed in oxygen-rich planetary nebulae. This strongly suggests that carbonaceous material is responsible for the ERE \citep{Witt+Vijh_2004}. In reflection nebulae, ERE is seen only when the exciting star has $T_{\rm eff}>10,000$\,K, hot enough to provide ample far-UV radiation \citep{Darbon+Perrin+Sivan_1999}. From the spatial distribution in IC59 and IC63, \citet{Lai+Witt+Crawford_2017} argue that ERE is excited by $11<h\nu<13.6$\,eV far-UV photons. Observed ERE intensities in reflection nebulae indicate overall photon conversion efficiencies (ERE photons emitted per UV photon absorbed) of $\lesssim 1\%$ \citep{Smith+Witt_2002}.

A number of authors have reported detection of the ERE from dust in Galactic cirrus clouds in the general ISM \citep{Guhathakurta+Tyson_1989,Szomoru+Guhathakurta_1998, Gordon+Witt+Friedmann_1998,Witt+etal_2008}. \citet{Gordon+Witt+Friedmann_1998} estimated the ERE emissivity to be

\begin{equation}
    \frac{\rm ERE~photons}{\rm H~atoms} = 5.65\times10^{-14} \frac{\rm photons}{\rm s~H~atom}
\end{equation}
with a required quantum yield $10\pm3\%$ if the ERE is excited by absorbed photons in the $2.25$--$13.6$\,eV range. While certain materials do indeed have high quantum efficiencies \citep[e.g., multilayer structures of SiO$_{0.9}$/SiO$_2$ luminesce at $\sim0.9\,\mu$m with a quantum yield $\sim45\%$;][]{Valenta+etal_2019}, an overall yield of 10\% would strongly constrain candidate grain materials.

Furthermore, if the ERE is actually primarily excited by 11--13.6\,eV photons, as concluded by \citet{Lai+Witt+Crawford_2017}, then the ERE intensities reported by \citet{Gordon+Witt+Friedmann_1998} would require an overall quantum yield approaching 100\%. This would require that (1) the ERE must originate from a major grain component, one accounting for a substantial fraction of the far-UV absorption, and (2) this component must have  a quantum efficiency of order 100\% for emitting an ERE photon following a FUV absorption. We are not aware of any candidate grain materials  that could meet this requirement while complying with elemental abundance constraints, and the observed extinction properties of interstellar dust.

On the other hand, measurement of the 4000--9000\,\AA\ spectrum of the diffuse Galactic light using SDSS blank sky spectra found that the shape of the diffuse light spectrum was consistent with the scattered light expected for standard grain models \citep{Brandt+Draine_2012}. \citet{Brandt+Draine_2012} estimated that no more than $\sim10\%$ of the dust-correlated diffuse light at $\sim6500\,$\AA\ could be ERE. This upper limit is inconsistent with the claimed detections toward individual cirrus clouds \citep{Guhathakurta+Tyson_1989,Szomoru+Guhathakurta_1998, Gordon+Witt+Friedmann_1998}. Additional observations will be needed to resolve this conflict.

We will assume that dust in both reflection nebulae and the general ISM produces ERE when illuminated by 11\,eV $\lesssim h\nu<13.6$\,eV photons, with an overall photon conversion efficiency $\sim1\%$ as seen in bright reflection nebulae. This conversion efficiency could either be the result of a low conversion efficiency for a major dust component or high conversion efficiency emission from a minor dust component (e.g., elements of the PAH population).  

In addition to the ERE, there is evidence for luminescence in the blue, peaking near $\sim3750\,$\AA, in the Red Rectangle \citep{Vijh+Witt+Gordon_2004} and in four reflection nebulae \citep{Vijh+Witt+Gordon_2005}. \citet{Vijh+Witt+Gordon_2004} suggested that the emission is fluorescence in small, neutral PAHs, containing 3--4 rings,  such as anthracene (C$_{14}$H$_{10}$) and pyrene (C$_{16}$H$_{10}$). It is not clear what abundance would be required to account for the blue luminescence.

\section{Polarized Emission}
\label{sec:ir_pol}
In this section we review observations of polarized infrared emission from interstellar dust and its connection to the observed polarized extinction.

\subsection{Infrared Emission}
\label{subsec:pir}

Just as aligned, aspherical grains polarize the starlight they absorb, the infrared emission from this same population of grains will be polarized.

The balloon-borne Archeops experiment \citep{Benoit_2002} provided a first look at polarized dust emission from the diffuse ISM in the Galactic plane. The 353\,GHz observations indicated polarization fractions of 4-5\%, with values exceeding 10\% in some clouds \citep{Benoit+etal_2004}, suggesting substantial alignment of the grains providing the submillimeter emission.

WMAP produced full-sky polarized intensity maps from 23--93\,GHz. Utilizing the final 9-year WMAP data, \citet{Bennett+etal_2013} found that the polarized dust emission $P_\nu$ in the WMAP bands is well-fit by a power law $P_{\nu} \propto \nu^{2+\beta}$ with $\beta = 1.44$. 

With polarimetric observations extending from 30 to 353\,GHz, the {\it Planck} satellite provided unprecedented constraints on the frequency dependence of the polarized emission. \citet{Planck_Int_XXII} found that the full-sky average of the polarized intensity of the dust emission from 100 to 353\,GHz is consistent with a modified blackbody having power law opacity $\kappa_\nu \propto \nu^\beta$ with $\beta = 1.59\pm0.02$ in contrast with $\beta = 1.51\pm0.01$ for total intensity over the same frequency range. This would imply a decrease in the polarization fraction between 353 and 70.4\,GHz with a significance greater than 3$\sigma$.

\begin{figure}
    \centering
        \includegraphics[width=\columnwidth]{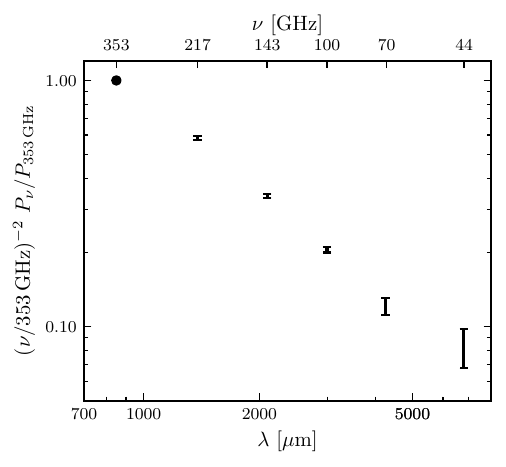}
    \caption{The submillimeter polarized dust SED as determined from the \texttt{SMICA} component separation algorithm on {\it Planck} polarization data \citep{Planck_2018_XI}. The normalization of this SED to the hydrogen column is discussed in Section~\ref{sec:pol_opt_ir}.} \label{fig:ir_pol} 
\end{figure}

Subsequently, \citet{Planck_2018_XI} employed the \texttt{SMICA} component separation algorithm to derive a global polarized dust SED. Despite making no assumptions on the parametric form of the dust SED, they found excellent agreement with a modified blackbody having $T_d = 19.6$\,K and $\beta = 1.53\pm0.02$. Following updates to the {\it Planck} photometric calibration, they revised the $\beta$ for total intensity to 1.48. With these changes, the $\beta$ determined for intensity and polarization are the same within $2\sigma$. In Figure~\ref{fig:ir_pol}, we plot the polarized dust SED of  \citet{Planck_2018_XI} and adopt it as a model constraint. We discuss the normalization of this SED to the hydrogen column in Section~\ref{sec:pol_opt_ir}.

\begin{deluxetable}{cc}
  \tablewidth{0pc}
      \tablecaption{Dust Polarization Fraction\label{table:pfrac}}
    \tablehead{\colhead{$\lambda$} & \colhead{\hspace{2cm}$p\left(\nu\right)/p_{353}$}\hspace{2cm}\\
    $\left[\mu{\rm m}\right]$ & }
    \startdata
    250 & $1.00\pm0.09$ \\
    350 & $1.06\pm0.11$ \\
    500 & $0.89\pm0.09$ \\
    850 & 1. \\
    1382 & $1.02\pm0.03$ \\
    2100 & $1.03\pm0.03$ \\
    3000 & $0.98\pm0.04$ \\
    4260 & $0.80\pm0.07$ \\
    6800 & $0.13\pm0.03$
    \enddata
    \tablecomments{The dust polarization fraction relative to 850\,$\mu$m (353\,GHz) as determined from BLASTPol observations in the Vela Molecular Ridge \citep{Ashton+etal_2018} and the {\it Planck} total and polarized dust SEDs \citep{Planck_Int_XVII, Planck_Int_XXII, Planck_2018_XI} compiled in Table~\ref{table:ir_sed}.}
\end{deluxetable}

\begin{figure}
    \centering
        \includegraphics[width=\columnwidth]{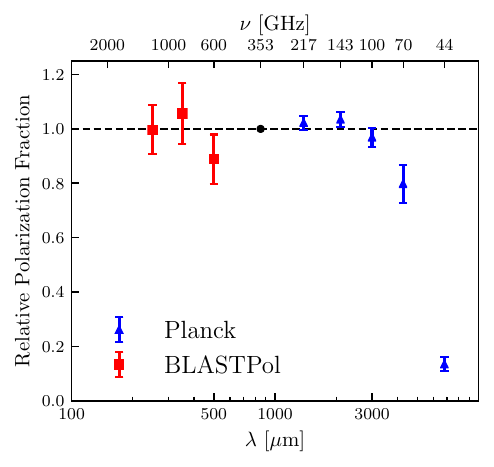}
    \caption{The polarization fraction of the dust emission relative to the 850\,$\mu$m (353\,GHz) polarization fraction as determined by BLASTPol and {\it Planck} observations. The BLASTPol data are from observations in the Vela Molecular Ridge \citep{Ashton+etal_2018} while the {\it Planck} data are based on the total and polarized dust SEDs \citep{Planck_Int_XVII,Planck_Int_XXII, Planck_2018_XI} compiled in Table~\ref{table:ir_sed}. Little wavelength dependence is observed except at the longest wavelengths where AME becomes a significant fraction of the total dust emission.} \label{fig:pfrac} 
\end{figure}

In addition to these large-scale observations of the diffuse ISM, polarimetric observations of dense clouds have also shed light on the FIR polarization properties of interstellar dust. In star-forming molecular clouds, the degree of polarization has been observed to fall from 60 to 350\,$\mu$m, then rise from 350 to 450\,$\mu$m \citep{Vaillancourt_2002, Vaillancourt+etal_2008}. This behavior can potentially be explained by correlated variations in the dust temperature and alignment efficiency in different regions along the line of sight, as might be expected in star-forming dense clouds.

However, BLASTPol observations of the Vela~C molecular cloud region and the Carina Nebula have revealed very little ($\lesssim 10\%$) evolution in the dust polarization fraction between 250 and 850\,$\mu$m \citep{Gandilo+etal_2016, Ashton+etal_2018, Shariff+etal_2019}. In particular, \citet{Ashton+etal_2018} studied translucent sightlines in the Vela Molecular Ridge, which is more likely to resemble diffuse sightlines than the observations at higher column densities. We present their determination of the dust polarization fraction, normalized to unity at 353\,GHz, in Table~\ref{table:pfrac} alongside the polarization fractions implied by the polarized dust SED of \citet{Planck_2018_XI} presented in Table~\ref{table:ir_sed}.

Taken together, the {\it Planck} and BLASTPol results suggest a roughly constant dust polarization fraction between 250\,$\mu$m and 3\,mm, as shown in Figure~\ref{fig:pfrac}.

Because very small nanoparticles are not expected to be aligned \citep{Draine+Hensley_2016}, polarization in the MIR dust emission features is generally not expected due to the small sizes of the grains able to emit at these wavelengths. However, a detection of polarization in the 11.3\,$\mu$m PAH feature has been reported in the nebula associated with the Herbig~Be star MWC~1080 \citep{Zhang+etal_2017}. If the polarization is indeed resulting from aligned PAHs, this may have implications for the theory of alignment of ultrasmall grains and thus predictions of AME polarization \citep{Draine+Hensley_2016,Hoang+Lazarian_2018}. However, it is not clear that either the dust properties or physical conditions in this region are likely to typify the diffuse ISM, and so we do not employ this result as a dust model constraint.

\subsection{Connection to Optical Polarization}
\label{sec:pol_opt_ir}
Because the same grains are believed to provide both polarized extinction in the optical and polarized emission in the infrared, it is expected that these quantities should be tightly related. Indeed, the polarization fraction of the 353\,GHz submillimeter emission \replaced{$p_S$}{$p_{353}$} divided by the $V$-band polarization per optical depth $p_V/\tau_V$ has a characteristic value between 4 and 5 over a range of column densities \citep[$N_{\rm H} \lesssim 5\times10^{21}\,$cm$^{-2}$, ][]{Planck_Int_XXI, Planck_2018_XII}. We adopt the best-fit value of 4.31 over diffuse sightlines \citep{Planck_2018_XII} as representative of dust in the diffuse ISM.

These relations between the polarized extinction and polarized emission from interstellar dust allow us to normalize the polarized dust SED derived by \citet{Planck_2018_XI} (see Figure~\ref{fig:ir_pol}) to the hydrogen column. First, \replaced{$p_S/\left(p_V/\tau_V\right)$}{$p_{353}/\left(p_V/\tau_V\right)$} $=  4.31$, $\left[p_V/E(B-V)\right]_{\rm max} = 0.13$, and $R_V = 3.1$ together imply a maximum  353\,GHz polarization fraction of 19.6\%, agreeing well with the observed maximum of \replaced{$p_S$}{$p_{353}$} $ = 22^{+3.5}_{-1.4}$\% \citep{Planck_2018_XII}. Applying this polarization fraction to the adopted 353\,GHz dust emission per H (see Table~\ref{table:ir_sed}) yields a maximum 353\,GHz polarized dust emission per H of $2.51\times10^{-28}$\,erg\,s$^{-1}$\,sr$^{-1}$\,H$^{-1}$. The polarized dust SED of \citet{Planck_2018_XI}, which is normalized to unity at 353\,GHz, can then be used to compute the maximum polarized dust emission per H at lower frequencies, as presented in Table~\ref{table:ir_sed}. We have color corrected all values, including corrections both at the observed frequency and at 353\,GHz, to obtain monochromatic spectral energy densities which can be compared directly to models.

\citet{Planck_Int_XXI} also introduced the ratio of the 353\,GHz polarized intensity to the V-band extinction, i.e, \replaced{$p_S$}{$P_{353}$}$/p_V$. They found a characteristic value of $5.4\pm0.5$\,MJy\,sr$^{-1}$ on translucent sightlines \added{integrated over the {\it Planck} 353\,GHz band}. \citet{Planck_2018_XII} extended this analysis to diffuse lines of sight, finding a characteristic ratio of $5.42\pm0.05$\,MJy\,sr$^{-1}$ with a systematic decrease to roughly 5\,MJy\,sr$^{-1}$ at the lowest ($\lesssim1\times10^{20}$\,cm$^{-2}$) column densities observed. This ratio is not independent of values we have already adopted:

\begin{equation}
\frac{P_{353}}{p_V} = \frac{1.086}{R_V} \frac{N_{\rm
                             H}}{E(B-V)} \frac{I_{353}}{N_{\rm H}}
                             \frac{p_{353}}{p_V/\tau_V} = 4.82\,{\rm
                               MJy}\,{\rm sr}^{-1}
~~~,
\end{equation}
\added{where here $I_{353}$ and $P_{353}$ are monochromatic. To compare with the \citet{Planck_2018_XII} value for the 353\,GHz band, we must multiply by the color correction factor 1.11 \citep{Planck_Int_XVII}, yielding 5.33\,MJy\,sr$^{-1}$ and good agreement.} \deleted{which is consistent with observations at low column densities.} We note, however, that the highly polarized region studied by \citet{Panopoulou+etal_2019} has $P_{353}/p_V = 4.1\pm0.1$\,MJy\,sr$^{-1}$, significantly lower than these values. Further study of this ratio and its variability across the sky is needed to understand this apparent discrepancy.

Performing a similar comparison between optical and submillimeter polarization in the Vela~C molecular cloud, \citet{Santos+etal_2017} related the \replaced{500\,$\mu$m}{600\,GHz (500\,$\mu$m)} polarization fraction \replaced{$p_{500}$}{$p_{600}$}, $I$ band polarized extinction $p_I$, and $V$ band total extinction, finding a characteristic \replaced{$p_{500}$}{$p_{600}$}$/\left(p_I/\tau_V\right) = 2.4\pm0.8$. For a typical Serkowski Law (see Section~\ref{sec:extpol_wav}), $p_V$ and $p_I$ differ by only about 10\%. If the FIR polarization fraction is relatively flat between \replaced{500 and 850\,$\mu$m}{353 and 600\,GHz} (see Figure~\ref{fig:pfrac}), then \replaced{$p_{500}$}{$p_{600}$}$/\left(p_I/\tau_V\right)$ should be approximately 10\% larger than \replaced{$p_S/\left(p_V/\tau_V\right)$}{$p_{353}/\left(p_V/\tau_V\right)$}, which has characteristic value 4.31 \citep{Planck_2018_XII}. This apparent discrepancy may be due to the very different environments sampled by these observations, but given the importance of this ratio in constraining models, further investigation is warranted.

\subsection{AME Polarization}
\label{subsec:ame_pol}
If the AME arises from aligned, aspherical grains, then it too will be polarized. However, searches for polarized AME have thus far yielded only upper limits at the $\sim1$\% level \citep{Dickinson+Peel+Vidal_2011,Macellari+etal_2011,GenovaSantos+etal_2015, Planck_2015_XXV,GenovaSantos+etal_2017}. See \citet{Dickinson+etal_2018} for a recent review.

To the extent that the smallest interstellar grains produce AME through rotational electric dipole radiation, the amount of AME polarization depends on how well these grains are able to align with the local magnetic field. The lack of polarization in the UV extinction curve (see Section~\ref{sec:extpol}) despite strong total extinction in the UV (see Section~\ref{subsec:ext_uv}) suggests that these grains are poorly aligned. However, \citet{Hoang+Lazarian+Martin_2013} demonstrated that if aligned PAHs were responsible for the claimed detections of polarization in the 2175\,\AA\ feature and also produced the AME, then the AME should be polarized at the $\lesssim 1\%$ level, near current upper limits. On the other hand, \citet{Draine+Hensley_2016} argued that quantization of the vibrational energy levels in ultrasmall grains leads to exponential suppression of their alignment, resulting in negligible AME polarization.

\section{Summary and Discussion}
\label{sec:summary}
\begin{figure*}
    \centering
       \includegraphics[width=\textwidth]{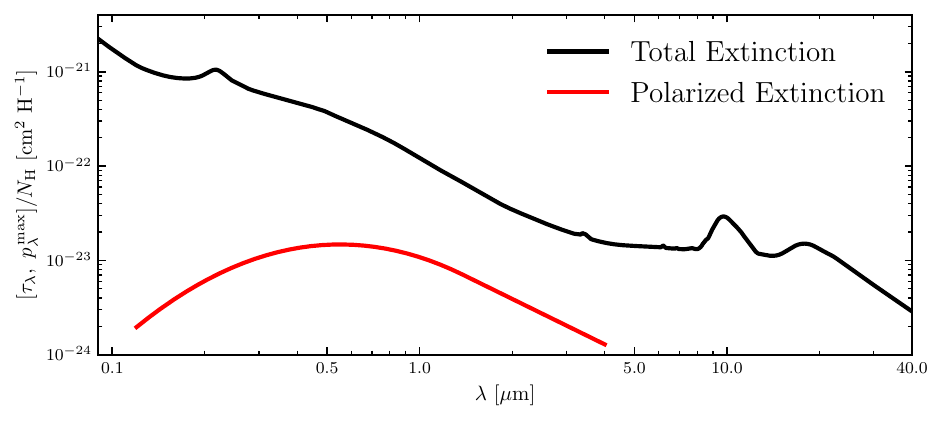}
       \includegraphics[width=\textwidth]{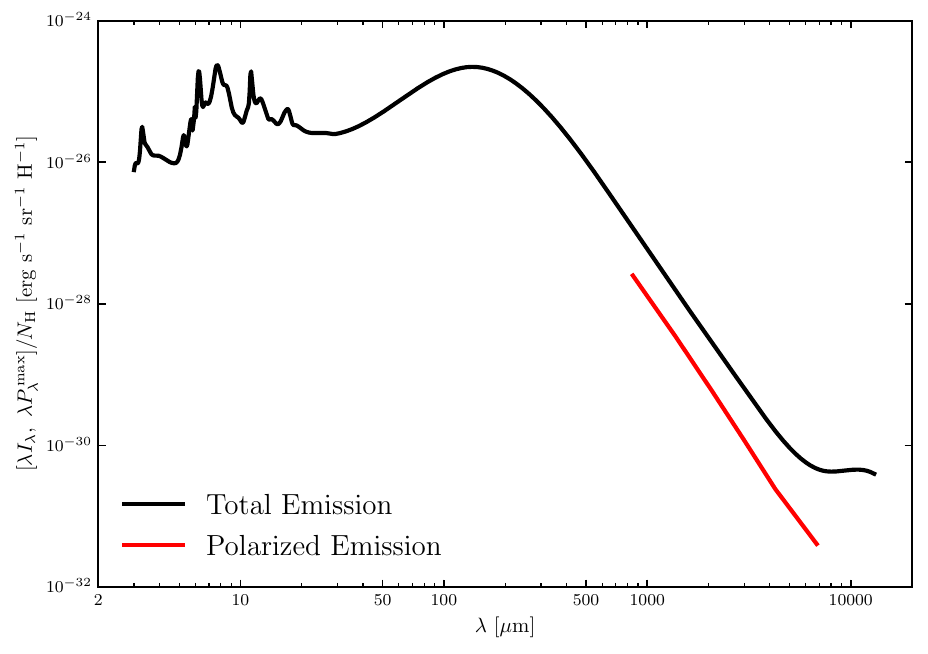}
    \caption{In the top panel, we plot our adopted constraints on the total (black) and polarized (red) extinction from dust in the diffuse ISM. In the bottom panel, we plot our adopted constraints on the total (black) and polarized (red) emission from interstellar dust. Note that for both polarized extinction and emission, we show the maximum level of polarization, corresponding the interstellar magnetic field lying in the plane of the sky. We have made use of the values in Table~\ref{table:constant_summary} where necessary to normalize the observational data to the hydrogen column. The data underlying the FIR emission constraints, including uncertainties, are presented in Table~\ref{table:ir_sed}. A summary of the adopted constraints is given in Section~\ref{sec:summary}. The extrapolation of the extinction curve to FIR wavelengths can be found in \citet{Hensley+Draine_2020}. The data behind this figure are available in tabular form from the journal.} \label{fig:constraints}
\end{figure*}

\begin{deluxetable*}{ccc}
  \tablewidth{0pc}
      \tablecaption{Adopted Values of Select Quantities for the
        Diffuse ISM\label{table:constant_summary}}
      \tablehead{\multicolumn{3}{c}{Reference Quantities}}
    \startdata
        Quantity & Value  &
      Reference \\ \hline
     $A( 5500\,\text{\AA})/E(4400\,\text{\AA}-5500\,\text{\AA})$ & 3.02 & \citet{Fitzpatrick+etal_2019} \\
    $A_H/A_{K_s}$ & 1.55 & \citet{Indebetouw+etal_2005} \\
    $N_H/E(B-V)$ & $8.8\times10^{21}$\,cm$^{-2}$\,mag$^{-1}$ &
    \citet{Lenz+Hensley+Dore_2017} \\
    $\left[p_V/E(B-V)\right]_{\rm max}$ & 0.13\,mag$^{-1}$ &
    \citet{Planck_2018_XII} \\
    $p_{353}/\left(p_V/\tau_V\right)$ & 4.31 & \citet{Planck_2018_XII} \\
    \hline \hline
    \multicolumn{3}{c}{Derived Quantities} \\ \hline
            Quantity & Value  & Reference\\ \hline
    $R_V$ & 3.1 & \citet{Fitzpatrick+etal_2019} \\
    $A_V/N_{\rm H}$ & $3.5\times10^{-22}$\,mag\,cm$^2$ & \\
    $p_{353}^{\rm max}$ & 19.6\% & \\
    $P_{353}/p_V$ & \replaced{4.8}{4.82}\,MJy\,sr$^{-1}$ &
    \enddata
    \added{\tablecomments{$P_{353}$ in the quantity $P_{353}/p_V$ is the monochromatic polarized intensity at 353\,GHz.}}
\end{deluxetable*}

Based on the foregoing discussion, we argue that the following data represent the current state of observations that constrain models of interstellar dust, and so a successful model of interstellar dust in the diffuse ISM should be measured against its consistency with these data. We also present a table of constants (Table~\ref{table:constant_summary}) based on observational data that enables the translation of these observables into constraints on the material properties of dust.

\begin{itemize}
\item {\bf Abundances}: Our adopted ISM abundances, given in Table~\ref{table:solid_abundance}, are based on Solar reference abundances \citep{Asplund+etal_2009,Scott+etal_2015b,Scott+etal_2015a} corrected for diffusion \citep{Turcotte+Wimmer-Schweingruber_2002} and with chemical enrichment \citep{Chiappini+Romano+Matteucci_2003, Bedell+etal_2018}. Solid phase abundances are then derived based on the gas phase abundances determined by \citet{Jenkins_2009} for sightlines with moderate depletion ($F_* = 0.5$).
\item {\bf Extinction}: We synthesize the extinction curves of \citet{Gordon+etal_2009} and \citet{Cardelli+Clayton+Mathis_1989} in the FUV, which we join to that of \citet{Fitzpatrick+etal_2019} in the UV through the optical. From 0.55 to 2.2\,$\mu$m, we employ the extinction curve of \citet{Schlafly+etal_2016} assuming $A_H/A_K = 1.55$ \citep{Indebetouw+etal_2005}. From 2.2 to 37\,$\mu$m, we adopt the MIR extinction curve derived by \citet{Hensley+Draine_2020} on the sightline toward Cyg\,OB2-12. Finally, we normalize this composite curve to the hydrogen column via $N_{\rm H}/E(B-V) =8.8\times10^{21}$\,cm$^{-2}$\,mag$^{-1}$ \citep{Lenz+Hensley+Dore_2017}.
\item {\bf Polarized Extinction}: Between 0.12 and 4\,$\mu$m, we join a Serkowski Law with parameters $K = 0.87$ and $\lambda_{\rm max} = 0.55$\,$\mu$m \citep{Whittet_2003} smoothly to a power law with index $\beta = 1.6$ in the IR \citep{Martin+etal_1992}. We normalize this curve to a maximum starlight polarization of $p_V/E(B-V) = 0.13$\,mag$^{-1}$ \citep{Planck_2018_XII, Panopoulou+etal_2019}.
\item {\bf Emission}: In the MIR, we adopt the {\it AKARI} and {\it Spitzer} spectrum of \replaced{the star-forming galaxy NGC\,5992 \citep{Brown+etal_2014}}{a sample of PAH-bright galaxies \citep{Lai+etal_2020}} between 3 and 12\,$\mu$m and the {\it Spitzer} IRS observations of the translucent cloud DCld 300.2-16.9 (B) \citep{Ingalls+etal_2011} between 6 and 38\,$\mu$m. The composite spectrum is scaled to the hydrogen column to match observations of diffuse Galactic emission in the DIRBE bands \citep{Dwek+etal_1997}. In the FIR, we adopt the $\ion{H}{i}$-correlated dust emission measured in the DIRBE and {\it Planck} bands with $\nu \ge 353$\,GHz \citet{Planck_Int_XVII}, and the 353\,GHz-correlated emission measured in the lower frequency {\it Planck} and WMAP bands by \citet{Planck_Int_XXII}.
\item {\bf Polarized Emission}: We adopt the frequency dependence of the polarized infrared emission determined by \citet{Planck_2018_XI} scaled to match the relation between polarized extinction and emission derived by \citet{Planck_2018_XII}.
\end{itemize}
These constraints are summarized visually in Figure~\ref{fig:constraints}, which illustrates the impressive breadth of our current knowledge, spanning a large dynamic range in wavelength, magnitudes of extinction, and intensity\replaced{,}{.} \replaced{and}{It also} highlights the most pressing needs for augmenting the state of art. We close by highlighting a few such future directions of key importance for dust modeling.

The spectroscopic features in extinction, emission, and polarization are the ``fingerprints'' of the specific materials that constitute interstellar grains, enabling determination of their chemical makeup. The Near InfraRed spectrograph \citep[NIRSpec, 0.6--5\,$\mu$m;][]{NIRSpec} and Mid-Infrared Instrument \citep[MIRI, 5--28\,$\mu$m;][]{MIRI} aboard the {\it James Webb Space Telescope} ({\it JWST}) will characterize the NIR and MIR spectroscopic dust features in unprecedented detail. Observing the full sky between 0.75 and 5\,$\mu$m with a resolving power of up to 130, SPHEREx \citep{SPHEREx} will enable mapping of the strength of dust absorption and emission features and thus probe their variation with location in the Galaxy. The high spectral resolution of the XRISM \citep{XRISM} and Athena \citep{Athena} X-ray observatories promises to reveal the mineralogical composition of interstellar grains in ways complementary to what can be gleaned from the infrared features.

As the 3.4\,$\mu$m complex has been observed on very few sightlines that might typify the diffuse ISM, a number of questions can be addressed by more sensitive observations. Is it indeed generic of the diffuse ISM that the 3.3\,$\mu$m aromatic feature is substantially broader in absorption than emission? To what extent does diamond-like carbon contribute emission and absorption in the 3.47\,$\mu$m feature? How does the 3.4\,$\mu$m profile change systematically with interstellar environment? 

The 6.2 and 7.7\,$\mu$m aromatic features have been observed in absorption, but on few sightlines. Detailed characterization of these features, particularly comparison of the emission and absorption profiles, will clarify which grains are the carriers of aromatic material in the ISM. The aromatic features at still longer wavelengths have not been observed in absorption, making them a compelling target for {\it JWST} and an important constraint on PAH models. 

While the aliphatic 6.85\,$\mu$m feature appears generic to the diffuse ISM on the basis of its detection in absorption toward Cyg\,OB2--12, the ubiquity of the 7.25\,$\mu$m methylene feature is less clear. Characterization of these aliphatic absorption features and their strengths relative to the aromatic features is a relatively unexplored window into the hydrocarbon chemistry of the ISM which {\it JWST} will enable. Likewise, the deuterated counterparts of both the aliphatic and aromatic features, inaccessible from the ground, will be accessible to {\it JWST} and SPHEREx in emission and absorption.

The sensitivity of MIRI will enable searches for as-yet undetected spectroscopic features and will characterize in greater detail those already observed. The silicate features can be probed for trace amounts of crystallinity, and the detection of crystalline forsterite can be verified on many more sightlines. Dedicated searches can be undertaken for the 11.2\,$\mu$m SiC feature and the 11.53\,$\mu$m feature from polycrystalline graphite, perhaps finally confirming or ruling out graphite as a major constituent of interstellar dust.

In the NIR, NIRSpec can characterize the many DIBs found longward of 600\,nm and perform sensitive searches for new ones. Likewise, the presence or absence of predicted features at 1.05 and 1.26\,$\mu$m from ionized PAHs can be strongly constrained.

While we anticipate advances in infrared spectroscopy, it is unfortunate that this is not the case for infrared spectropolarimetry. Polarimetry is a powerful complementary constraint on the properties of interstellar dust, particularly given the dichotomy observed in polarization between carbonaceous and silicate features. Additionally, the profiles of the spectroscopic features in extinction and polarization generically differ because each depends differently on the optical constants, and so measurement of both strongly constrains grain material properties. Additional spectropolarimetric measurements of the 9.7 and 18\,$\mu$m silicate features and the 3.4\,$\mu$m carbonaceous feature are desperately needed. In addition, the continuum polarization between 4--8\,$\mu$m is poorly determined. Unfortunately, we are unaware of any operational facilities, nor of any planned ones, capable of spectropolarimetry or even broadband polarimetry between 3 and 8\,$\mu$m. However, new polarimetric measurements of the 9.7\,$\mu$m silicate feature are possible with CanariCam \citep{CanariCam}.

Stellar optical polarimetry, on the other hand, will be pushed to high latitude, diffuse sightlines in the 2020s with the PASIPHAE survey \citep{PASIPHAE}. With a many-fold expansion of stellar polarization catalogues, new insights will be gained in the variations in the polarized extinction curve throughout the Galaxy, including its connection to polarized infrared emission.

Because of the role of dust polarization in mapping magnetic fields and as a contaminant for Cosmic Microwave Background (CMB) polarization science, the prospects are better for studies of polarized emission. Of critical importance from the perspective of dust modeling is extending coverage of the polarized dust SED to higher frequencies on sightlines that might typify the diffuse ISM. Measuring the wavelength-dependence of polarization near the peak of the dust SED will allow the contributions from different dust populations to be more efficiently disentangled. At even shorter wavelengths, we expect emission to be dominated by smaller, unaligned grains. While such measurements are already possible on dense sightlines using instruments like HAWC+ aboard SOFIA \citep{HAWC+}, the greater sensitivity afforded by upcoming facilities like CCAT-prime \citep[$260 \leq \nu/{\rm GHz} \leq 860$;][]{CCAT-prime} is required to access diffuse sightlines. However, we are unaware of upcoming facilities that can perform polarimetry on the Wien side of the dust SED along diffuse sightlines.

Particularly given the uncertainties in the level of polarization in the AME and the abundance of material able to emit microwave magnetic dipole radiation, extension of the determination of the polarized dust SED to lower frequencies is also of great interest. Such measurements will be made by upcoming CMB experiments such as the Simons Observatory \citep{SimonsObservatory}, LiteBIRD \citep{LiteBIRD}, and CMB-S4 \citep{CMB-S4}, all of which have the sensitivity to characterize dust emission on the diffuse, high latitude sightlines of greatest interest to this work.

These directions are but a few avenues to be explored with the wealth of upcoming data and are not intended to be exhaustive. As we emphasize in this work, dust modeling should be informed by the full range of optical phenomena associated with interstellar \replaced{grains,}{dust.} \replaced{and by}{By} combining the insights gleaned from a variety of observations across the electromagnetic spectrum, we can paint the clearest picture possible of the nature of interstellar grains.

\acknowledgments
{We are grateful to many stimulating conversations that informed this work over its long completion. We thank in particular Megan Bedell, \added{Simone Bianchi,} Tuhin Ghosh, Vincent Guillet, \deleted{Jim Ingalls,} Ed Jenkins, Eddie Schlafly, \added{Adolf Witt,} and Chris Wright for sharing their expertise. \added{We thank Jim Ingalls for providing the {\it Spitzer} IRS spectrum of DCld 300.2-16.9, and Thomas Lai and JD Smith for providing their ``1C'' PAH spectrum in advance of publication. We thank the anonymous referee for helpful comments.} This research was carried out in part at the Jet Propulsion Laboratory, California Institute of Technology, under a contract with the National Aeronautics and Space Administration. This work was supported in part by NSF grants AST-1408723 and AST-1908123. BH acknowledges support from the National Science Foundation Graduate Research Fellowship under Grant No. DGE-0646086 during the earliest stages of this work.}
\software{Astropy \citep{Astropy,Astropy_2}, Matplotlib \citep{Matplotlib}, NumPy \citep{NumPy}, SciPy \citep{SciPy}}

\bibliography{mybib}
\listofchanges
\end{document}